\documentclass[a4paper,11pt]{article}
\usepackage{thesis}
\usepackage{natbib}
\usepackage{graphicx}
\usepackage{transparent}
\usepackage[font={singlespacing,small,it}]{caption}
\begin{document}
\font\myfont=cmr12 at 15pt
\title{{\myfont An innovation diffusion model of a local electricity network that is influenced by internal and external factors}}
\date{}
\author{Laura Hattam and Danica Vukadinovi\'c Greetham\footnote{Department of Mathematics and Statistics, University of Reading, UK}}
\maketitle
\abstract{\citet{hay77} derived a nonlinear differential equation to determine the spread of innovations within a social network across space and time. This model depends upon the imitators and the innovators within the social system, where the imitators respond to internal influences, whilst the innovators react to external factors. Here, this differential equation is applied to simulate the uptake of a low-carbon technology (LCT) within a local electricity network situated in the UK. This network comprises of many households that are assigned to certain feeders. Firstly, travelling wave solutions of Haynes' model are used to predict adoption times as a function of the imitation and innovation influences. Then, the grid that represents the electricity network is created so that the finite element method (FEM) can be implemented. Next, innovation diffusion is modelled with Haynes' equation and the FEM, where varying magnitudes of the internal and external pressures are imposed. Consequently, the impact of these model parameters is examined. Moreover, LCT adoption trajectories at fixed feeder locations are calculated, which give a macroscopic understanding of the uptake behaviour at specific network sites. Lastly, the adoption of LCTs at a household level is discussed.}
\section{Introduction}
\citet{rog62} proposed that over time, the diffusion of a new technology through a social network resulted from the social interactions of its members. \citet{bas69} built upon this work by applying a nonlinear ordinary differential equation to predict innovation diffusion as a function of time, which contained parameters associated with the `innovators' and `imitators' within the network. From this, the `S-curve' solution that predicts the number of adopters with time was derived. Later, \citet{hay77} extended the Bass model to also include spatial dependence, which is of the form
\beq
\frac{\partial N}{\partial t}=\left(p+q N\right)\left(1-N\right)+ k\left(\frac{\partial^2 N}{\partial x^2}+ \frac{\partial^2 N}{\partial y^2}\right),\label{maine}
\eeq
where $x$ and $y$ are spatial variables, $t$ is time, $N(x,y,t)\in [0,1]$ represents the percentage of innovation adopters at the location $(x,y)$ and time $t$, $k$ is the diffusion coefficient, and $p$ and $q$ are the innovation and imitation coefficients respectively. The innovation parameter relates to external influences, such as advertising and the imitation coefficient corresponds to internal effects, like word-of-mouth. Here, (\ref{maine}) is applied to determine the uptake of low-carbon technologies (LCTs) within a sample area. After some manipulation of (\ref{maine}), it becomes the Fisher equation, which was used by \citet{fis37} to model the spread of a mutant gene.

\citet{mor70} described the innovation adoption process as a wave-like phenomenon propagating in space and time, using such physical examples as the implementation of tuberculosis controls within agricultural areas to demonstrate this. Therefore, known travelling wave solutions of the Fisher equation could be used to predict the spread of innovations across certain geographies. \citet{tys00} derived travelling wave solutions to the Fisher equation over two spatial dimensions. In particular, they defined the well-known Fisher travelling wave identified by \citet{abl79} in terms of two spatial variables. This solution is of the same shape as the S-curve, although it varies with space and time, and is used here to predict the adoption behaviour.

\citet{shi09} and \citet{kar16} used differential equations to model the spread of LCTs. They both studied geographic areas with complicated geometries and therefore, the finite element method (FEM) was utilised. Our area of interest is also complex as it is a local electricity network comprised of $249$ feeders and so the FEM will be applied here. The aim is to obtain S-curves at the feeder locations, which will be useful measures for electricity network planners. Currently, accurate values for the parameters $p$ and $q$ for the area under examination are unknown. However, over the coming years, further information about the uptake behaviour throughout the UK will presumably become available. As a result, $p$ and $q$ can then be obtained by fitting the data to an S-curve. Until then, the implications of changing $p$ and $q$ can be studied. Hence, a variety of scenarios can be explored with differing levels of external and internal pressures.

More specifically, in Section 2, a travelling wave solution is derived that is dependent upon the equation parameters $p$ and $q$. In addition, the variation of these parameters and the effect this will have on the adoption curves is investigated. Next, in Section 3, the geographic area of interest is outlined, as well, so to apply the FEM, the grid is constructed. Then, in Section 4, (\ref{maine}) is applied to this grid to simulate the spread of innovations and the results are depicted. Furthermore, innovation uptake at a household level is explored.

\section{The Model\label{mod}}
The innovation diffusion model proposed by \citet{hay77} is now transformed into the Fisher equation. Then, the analysis related to the Fisher equation is used to predict adoption times as a function of $p$ and $q$.

\begin{figure}
\begin{center}
\includegraphics[width=3.2in,height=2.4in]{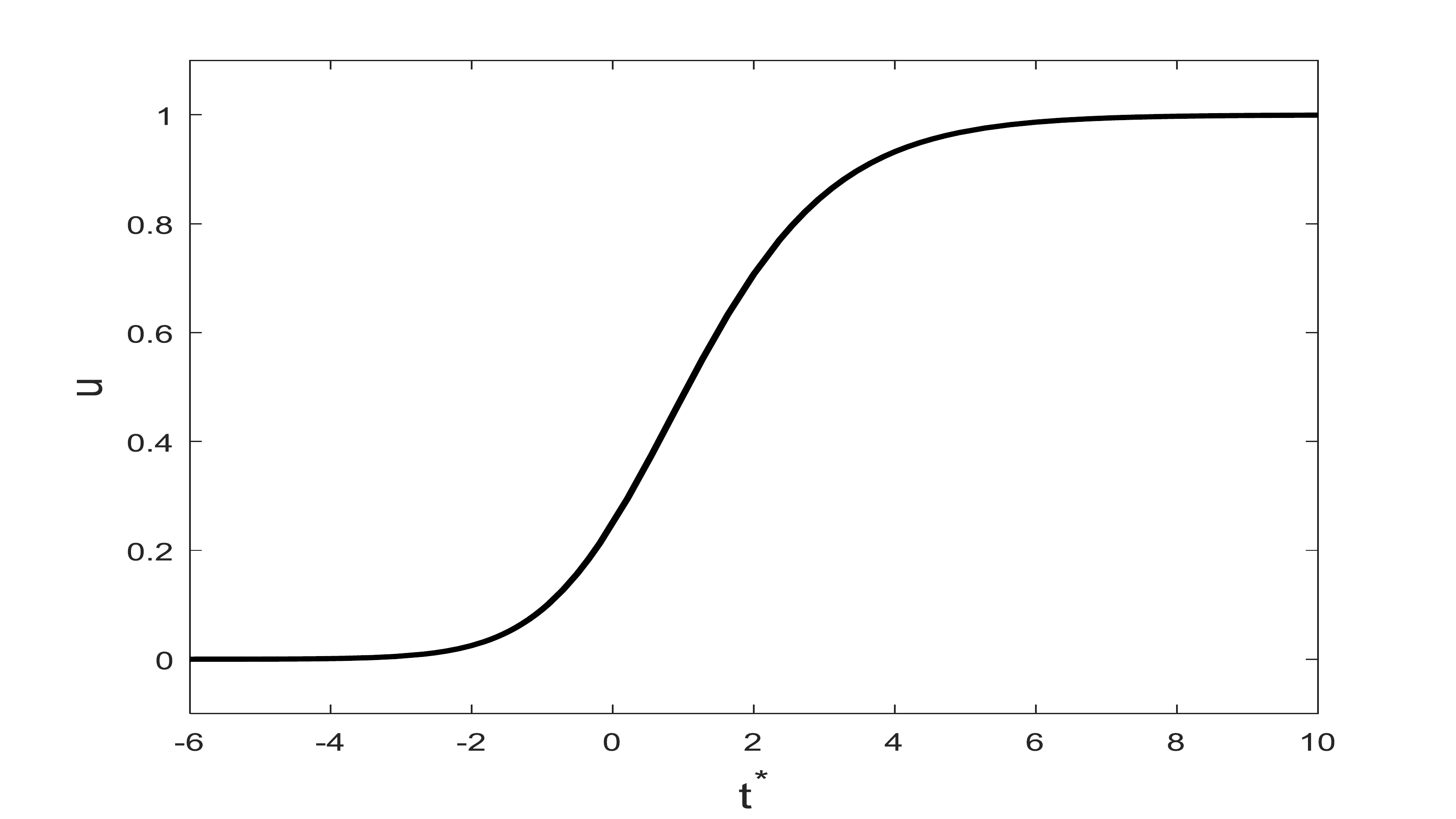}
\includegraphics[width=3.2in,height=2.4in]{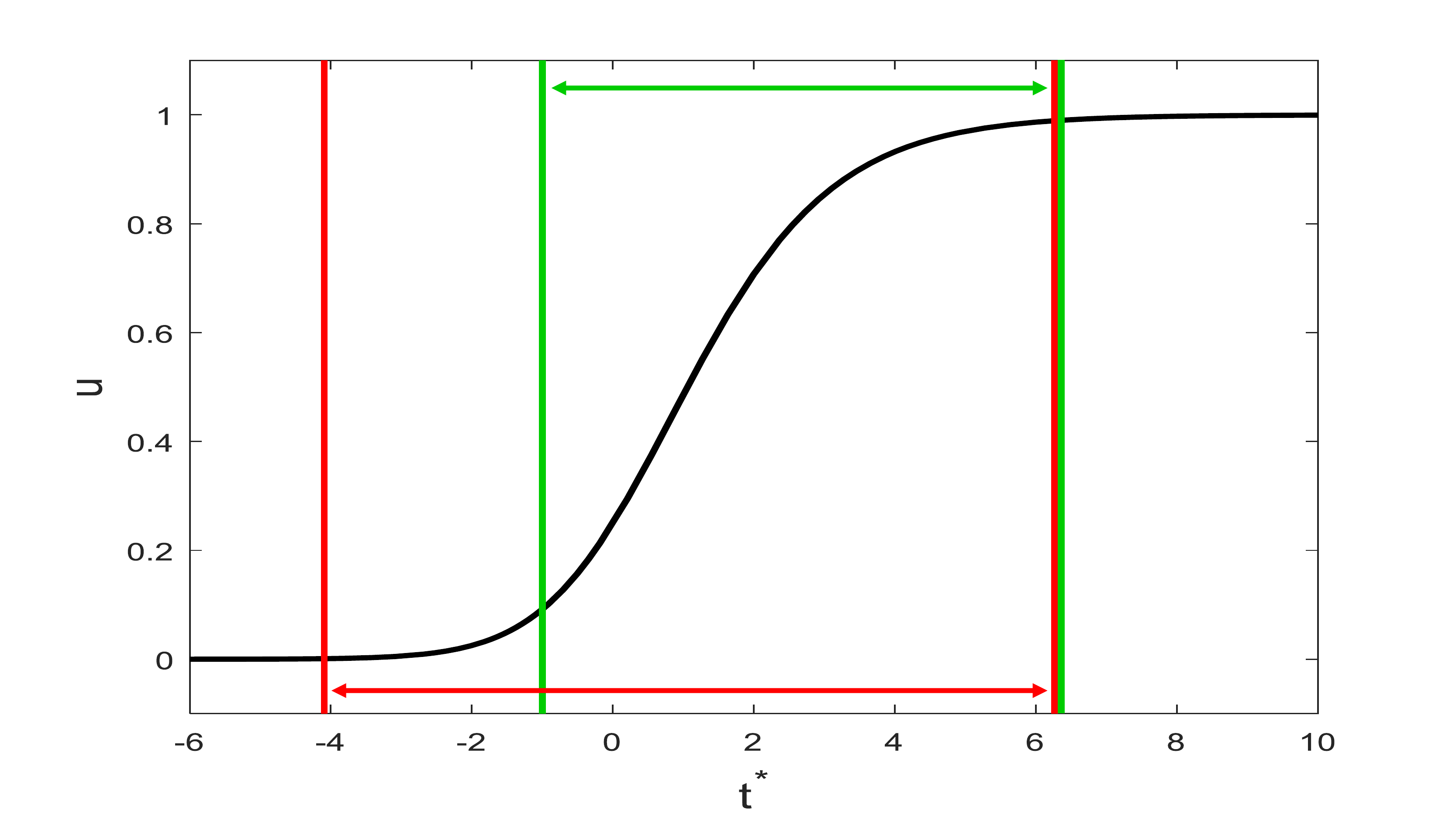}\\
\includegraphics[width=3.2in,height=2.4in]{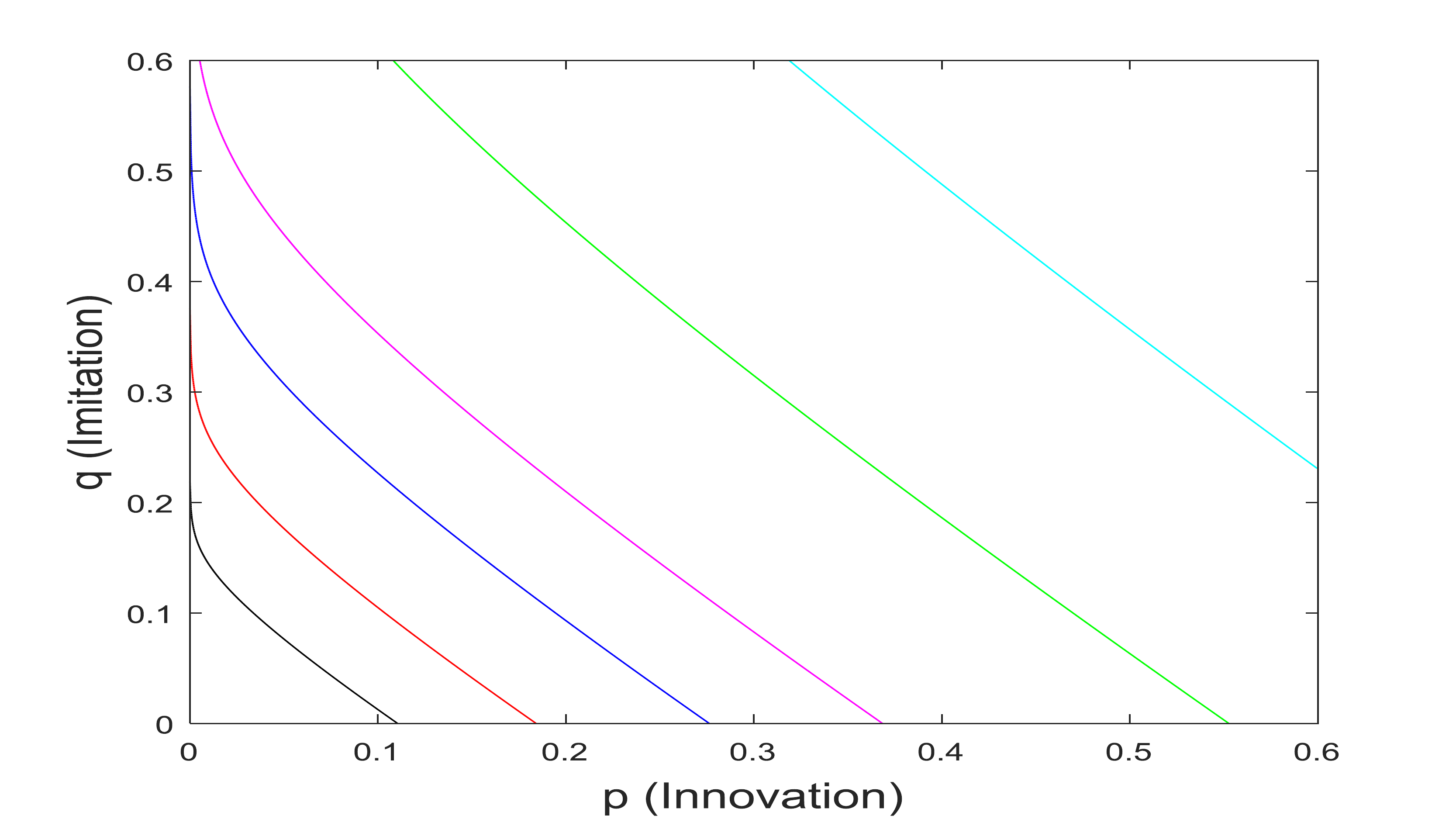}
\includegraphics[width=3.2in,height=2.4in]{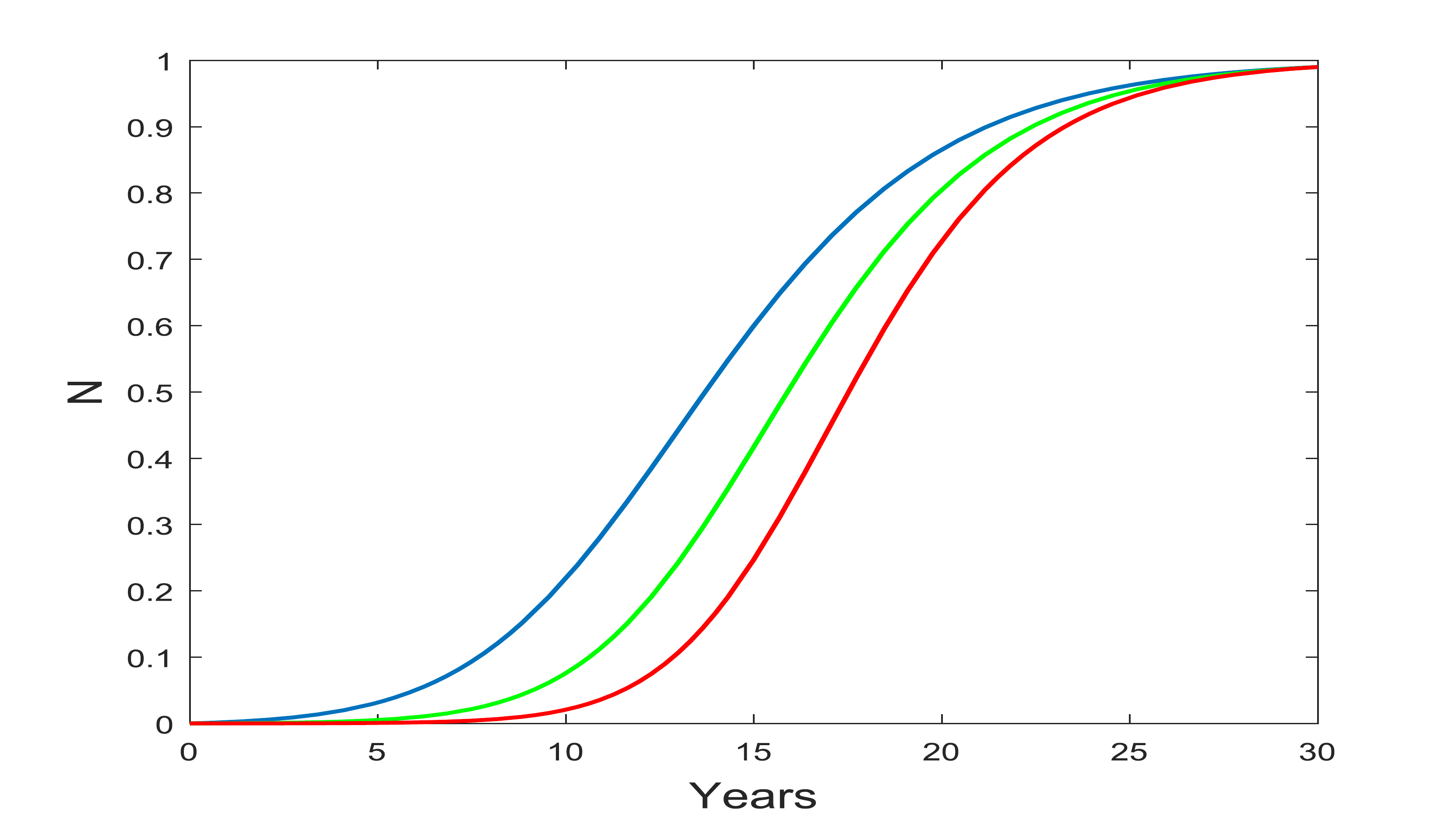}
\caption{Top Left: Plot of $u$ defined by (\ref{travwavefx}) when $\tilde{z}=0$. Top Right: Solution window for $p=0.0001, q=0.1$ (red) and $p=0.01, q=0.1$ (green). Bottom Left: Time required to achieve at least $99\%$ uptake everywhere as a function of $p$ (innovation coefficient) and $q$ (imitation coefficient), refer to (\ref{Delt}); black: $\Delta t=50$ years, red: $\Delta t=30$ years, blue: $\Delta t=20$ years, purple: $\Delta t=15$ years, green: $\Delta t=10$ years, light blue: $\Delta t=7$ years. Bottom Right: Innovation uptake curves for $p=0.001$, $q=0.325$ (blue), $p=0.0001$, $q=0.376$ (green) and $p=0.00001$, $q=0.425$ (red) defined by (\ref{travwaveN}).
\label{uplot}}
\end{center}
\end{figure}

To begin, let $t^{*}=(p+q)t$, $x^{*}=\sqrt{\frac{p+q}{k}}x$, $y^{*}=\sqrt{\frac{p+q}{k}}y$ and $N=(1+p/q)u-p/q$. As a result, (\ref{maine}) becomes
\beq
\frac{\partial u}{\partial t^{*}}=u\left(1-u\right)+ \frac{\partial^2 u}{\partial x^{*2}}+ \frac{\partial^2 u}{\partial y^{*2}}.\label{Fishe}
\eeq
This is the Fisher equation in two spatial dimensions. 
\citet{abl79} found an exact travelling wave solution to (\ref{Fishe}), which is
\beq
u=\frac{1}{\left(1+\exp\left(\frac{1}{\sqrt{6}}\left(x^{*}-ct^{*}\right)\right)\right)^2},
\eeq
where $c=\frac{5}{\sqrt{6}}$ is the wave speed. This wave propagates along the $x^{*}$-axis ($u_{y^{*}y^{*}}=0$) and is monotonically increasing with $t^*$, at fixed $x^*$. Then by rotating the frame of reference, \citet{tys00} identified the family of solutions
\beq
u=\frac{1}{\left(1+\exp\left(\frac{1}{\sqrt{6}}\left(x^{*}\sin\theta+ y^{*}\cos\theta-ct^{*}\right)\right)\right)^2},\label{travwave}
\eeq
such that the waves propagate at some angle $\theta$ from the $x^{*}$-axis. So, if our focus is how (\ref{travwave}) evolves with increasing $t^{*}$ at a fixed location, then let us set
\beq
u=\frac{1}{\left(1+\exp\left(\frac{1}{\sqrt{6}}\left(\tilde{z}-ct^{*}\right)\right)\right)^2},\label{travwavefx}
\eeq
where $\tilde{z}$ is some constant. This solution is plotted in Figure \ref{uplot}, top left with $\tilde{z}=0$ ($\tilde{z}\ne 0$ will simply shift the solution right or left).

Re-writing (\ref{travwave}) in terms of the adopter ratio, the following travelling wave solution to (\ref{maine}) is found
\beq
N=\left(1+\frac{p}{q}\right)\frac{1}{\left(1+\exp\left(\frac{1}{\sqrt{6}}\left(\sqrt{\frac{p+q}{k}}x\sin\theta+ \sqrt{\frac{p+q}{k}}y\cos\theta-c(p+q)t\right)\right)\right)^2}-\frac{p}{q}.\label{travwaveN}
\eeq
This solution has the wave speed $c\sqrt{k(p+q)}$ in $(x,y,t)$ space and therefore, when $p$ and $q$ are small, progression is slow. Since varying $k$ simply re-scales the spatial domain $(x^*,y^*)$, as well, given that the effects of $p$ and $q$ on the adoption times are the main focus here, we set $k=1$ without any loss of generality.

Now suppose we are interested in the time taken for N to increase to $0.99$, at a fixed location and beginning at zero, then, it is necessary to calculate
 \beq
 \Delta t=t_2-t_1=\frac{6}{5(p+q)}\log\left|\frac{\sqrt{q/p+1}-1}{\sqrt{\frac{1+p/q}{0.99+p/q}}-1}\right|,\label{Delt}
 \eeq
where $N(t_1)=0$ and $N(t_2)=0.99$. Therefore, specifying $p$ and $q$ determines $\Delta t$, the time domain of interest.

Next, let $u(t^{*}_1)=u_1=1/((q/p)+1)$ and $u(t^{*}_2)=u_2=(0.99+(p/q))/(1+(p/q))$ when $N=0$ and $N=0.99$ respectively, where $\Delta t^{*}=t^{*}_2-t^{*}_1=(p+q)\Delta t$ and the solution window is defined as $u\in[u_1,u_2]$. Therefore, varying $p$ and $q$ changes the solution domain. As an example, see Figure \ref{uplot}, top right, where the solution windows for the two cases $p=0.0001,q=0.1$ and $p=0.01,q=0.1$ are depicted. This figure demonstrates that if $p/q$ is very small such that the imitation (internal) effect dominates, the estimated uptake is sluggish in the beginning and further time is needed before there is any evident rise in the overall adoption ratio. This behaviour is due to uptake being initially very localised. In contrast, when $p/q$ is greater and the innovation coefficient is amplified, adoptions are almost immediate. Uptake is now no longer at a local level but the process instead occurs more uniformly across the entire population. 
In Section 4, the influences of $p$ and $q$ on the local electricity network are further explored.

Due to (\ref{Delt}), a solution parameter space can be defined in terms of $p$, $q$ and $\Delta t$, which is shown in Figure \ref{uplot} bottom left. Here, $\Delta t= 30$ years is used as a hypothetical case, which means that overall at least $99\%$ of our sample population will have adopted the technology in $30$ years. We will study three scenarios corresponding to $\Delta t=30$ years that differ in their levels of internal and external pressures. These are $(a)$ $p=0.001$, $q=0.325$, $(b)$ $p=0.0001$, $q=0.376$ and $(c)$ $p=0.00001$, $q=0.425$. The adoption trajectories for these three cases are shown in Figure \ref{uplot}, bottom right, defined by (\ref{travwaveN}). Note that these curves all converge as $t\rightarrow 30$ years.

\section{The Grid}
An analysis of a local electricity network in Bracknell, UK is now undertaken, which comprises of $9484$ households and $249$ feeders. Every household is connected to a certain feeder, where electricity is supplied to a home along its assigned feeder. Therefore, these assignments represent spatial clusters of households. Here, (\ref{maine}) is applied to model innovation diffusion amongst these feeders using the FEM, where the imitation and innovation coefficients are varied. As a result, we can assess the effects of internal and external pressures at a local level. More specifically, adoption trajectories will be determined for each feeder, which is the expected uptake at a fixed location as a function of time. These projections are important measures for network planners as they indicate the potential, additional demand at specific network sites. If extra load along a feeder is not accounted for at the design stage, then issues such as localised network damage or disconnections can result when subjected to periods of high demand.

We are concerned with the uptake of LCTs, including electric vehicles (EVs) and photovoltaics (PVs) since these are expected to be employed throughout the UK and to have a significant impact upon the existing electricity network.  The spread of LCTs across a geographical area can be modelled with (\ref{maine}). As an example, see \citet{shi09}, where the uptake of hybrid cars within Japan was predicted using (\ref{maine}).

\begin{figure}
\begin{center}
\includegraphics[width=3.2in,height=2.4in]{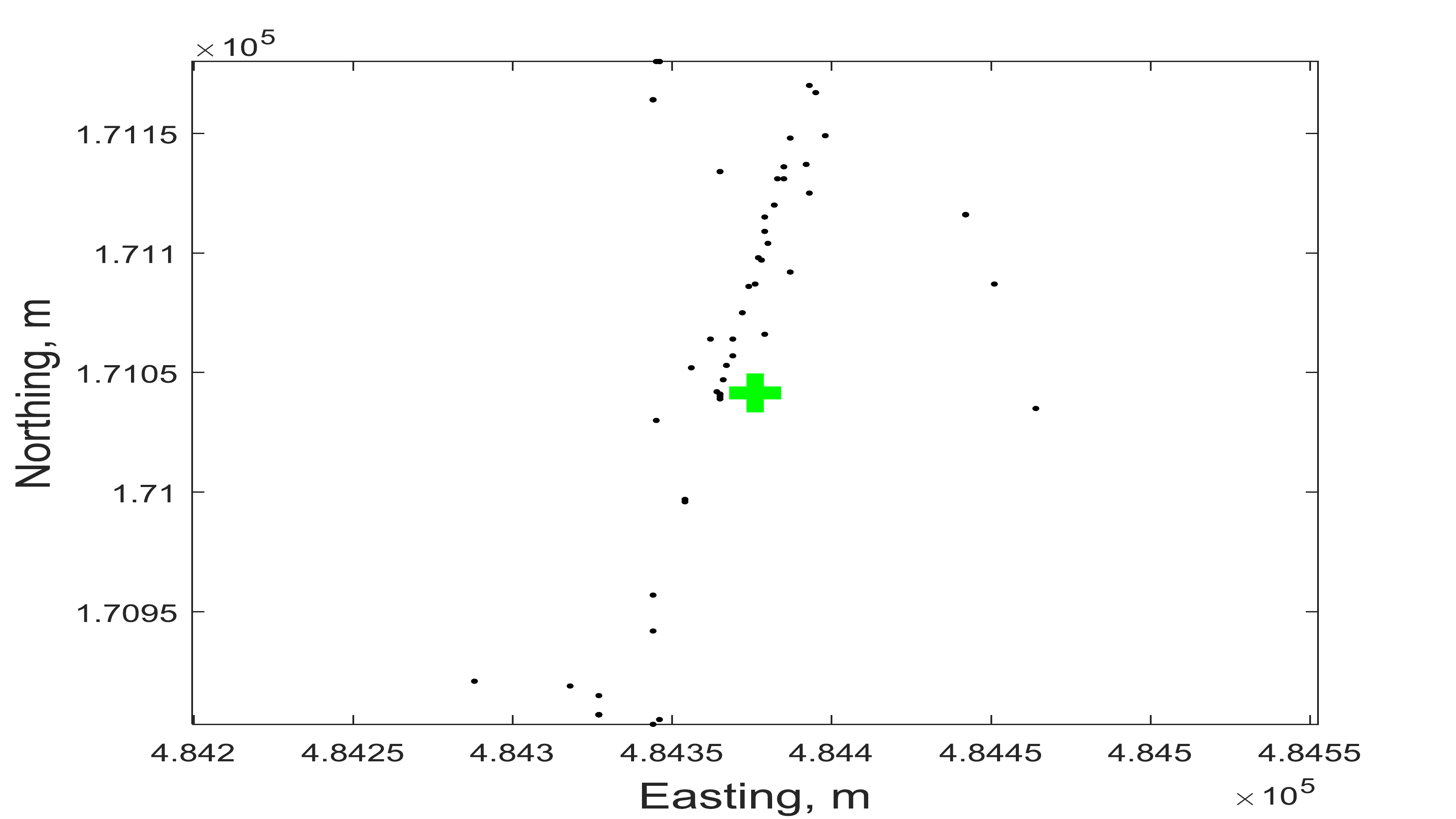}
\includegraphics[width=3.2in,height=2.4in]{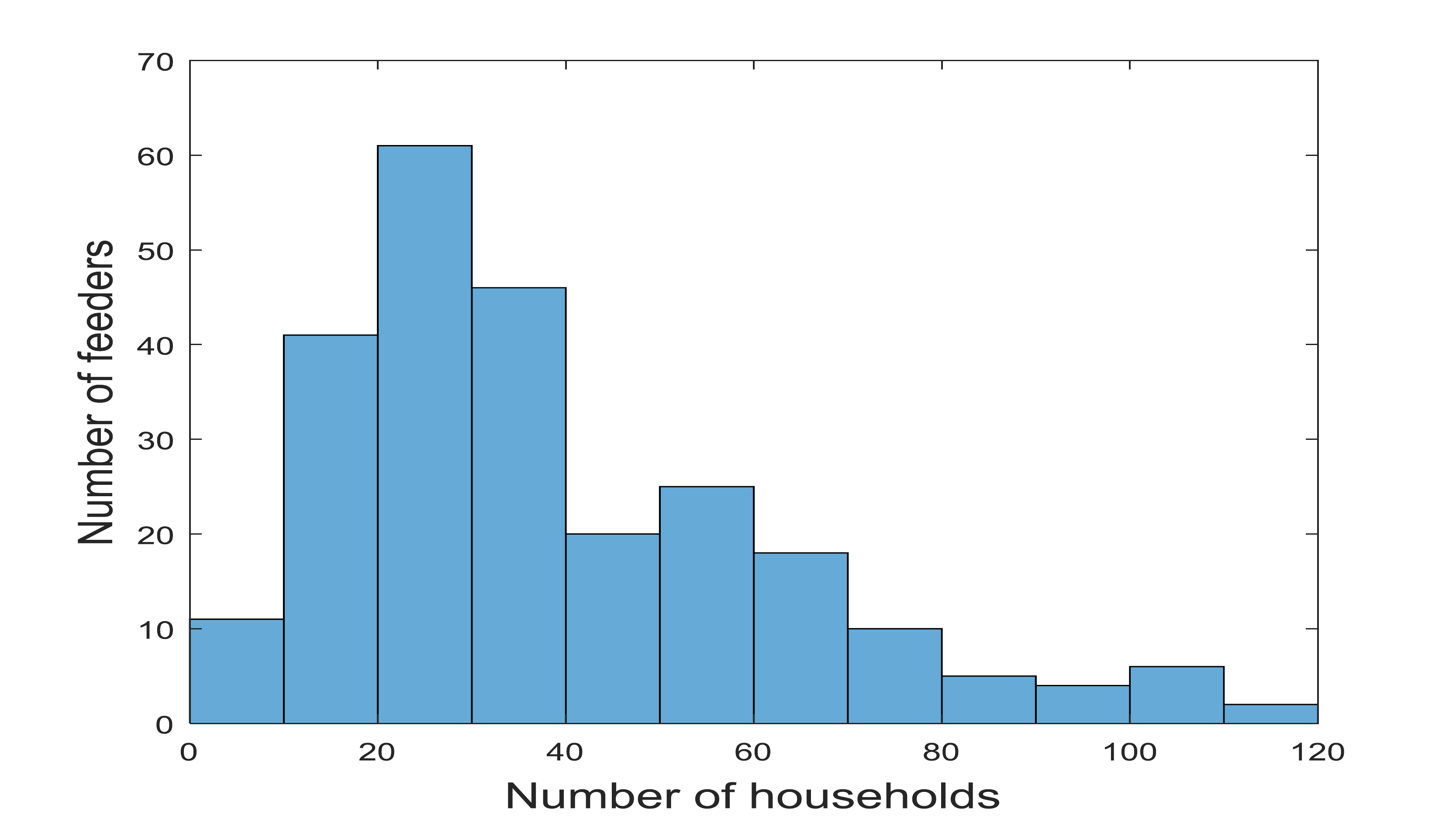}
\caption{Left: Determining the feeder co-ordinates (green cross) for a feeder with $84$ households (black dots). Right: The distribution of households amongst the $249$ feeders.\label{midpoint}}
\end{center}
\end{figure}

Firstly, our finite element grid needs to be generated. The feeder locations in $(x,y)$ space are used to determine this, where the units for $x$ and $y$ are easting (meters) and northing (meters) respectively. The $(x,y)$ co-ordinate of a feeder corresponds to the midpoint of all its households in the $x$ and $y$ directions. In Figure \ref{midpoint}, left, an example is given of this calculation for a feeder with $82$ households. It should be noted that the number of households assigned to each feeder varies somewhat. Figure \ref{midpoint}, right demonstrates the distribution of households amongst the $249$ feeders.

\begin{figure}
\begin{center}
\includegraphics[width=3.2in,height=2.4in]{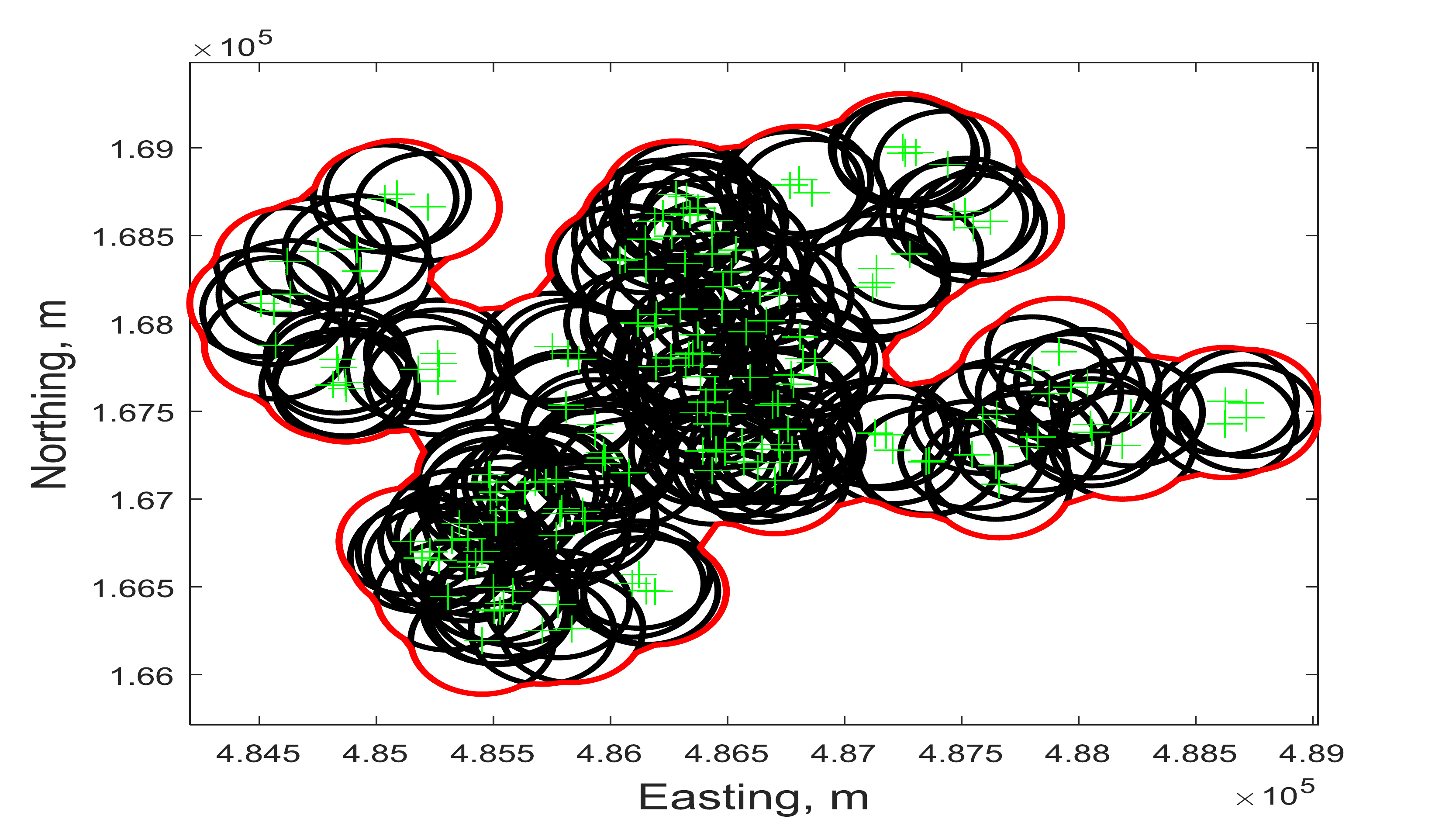}
\includegraphics[width=3.2in,height=2.4in]{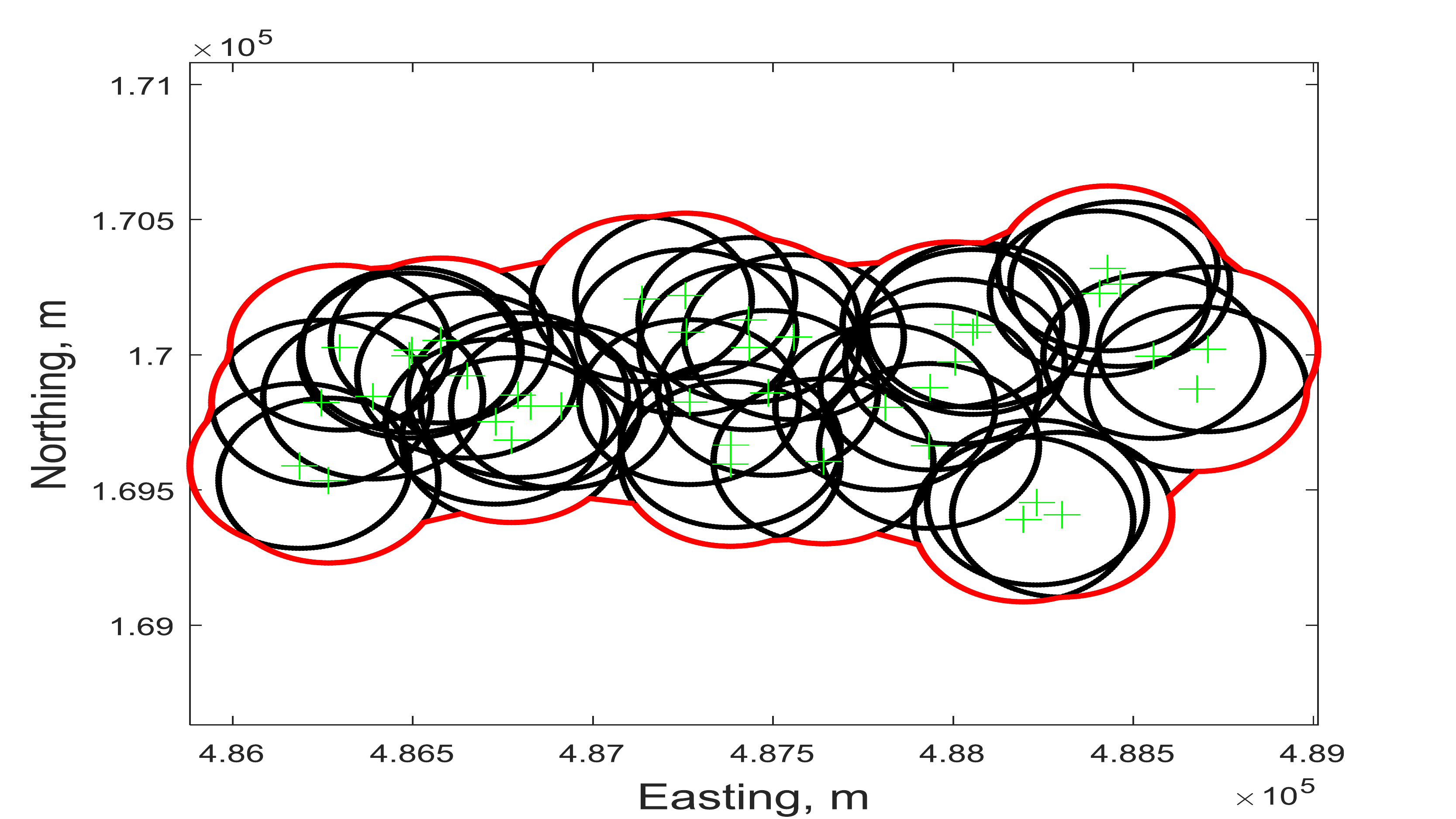}\\
\includegraphics[width=3.2in,height=2.4in]{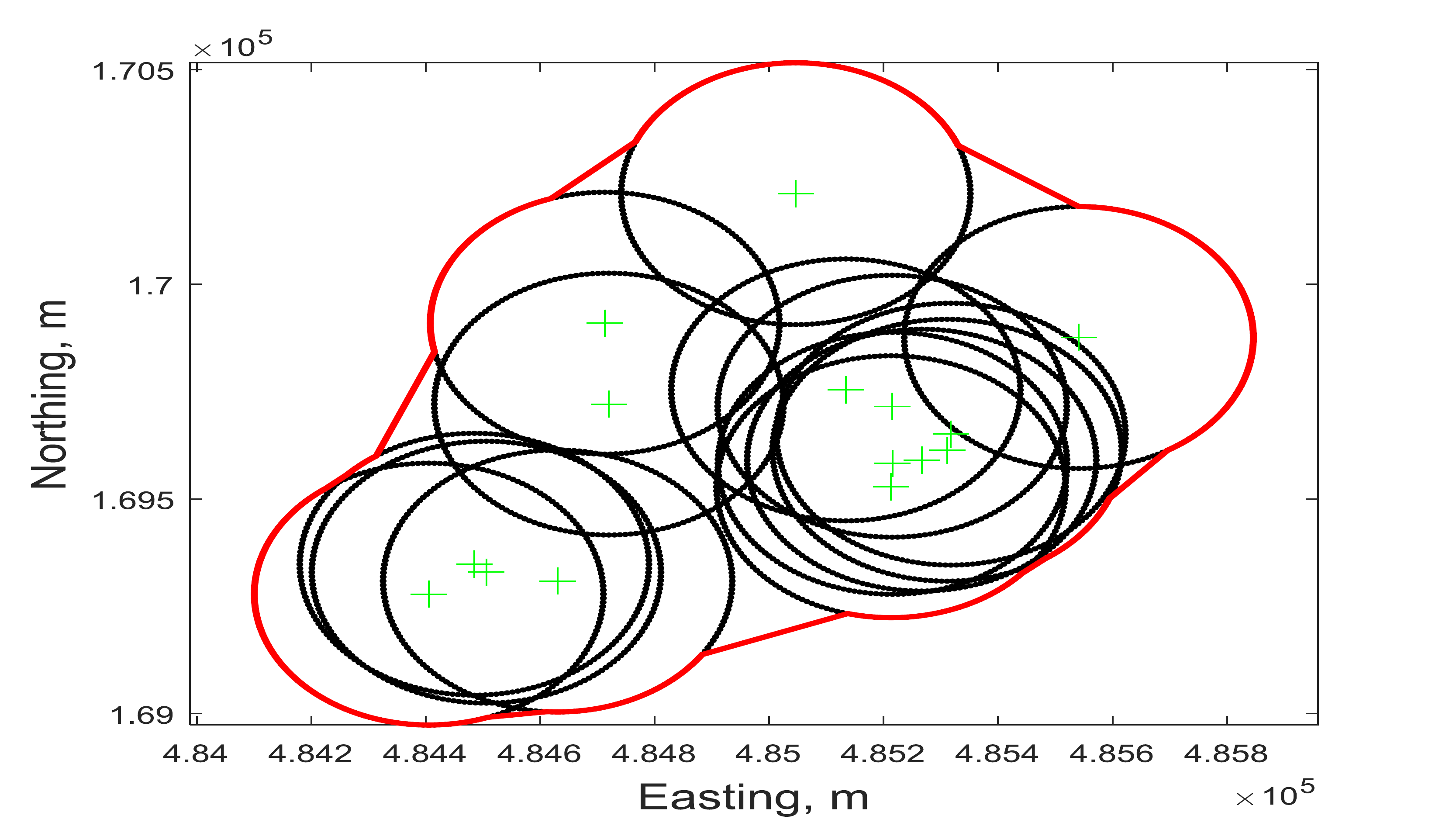}
\includegraphics[width=3.2in,height=2.4in]{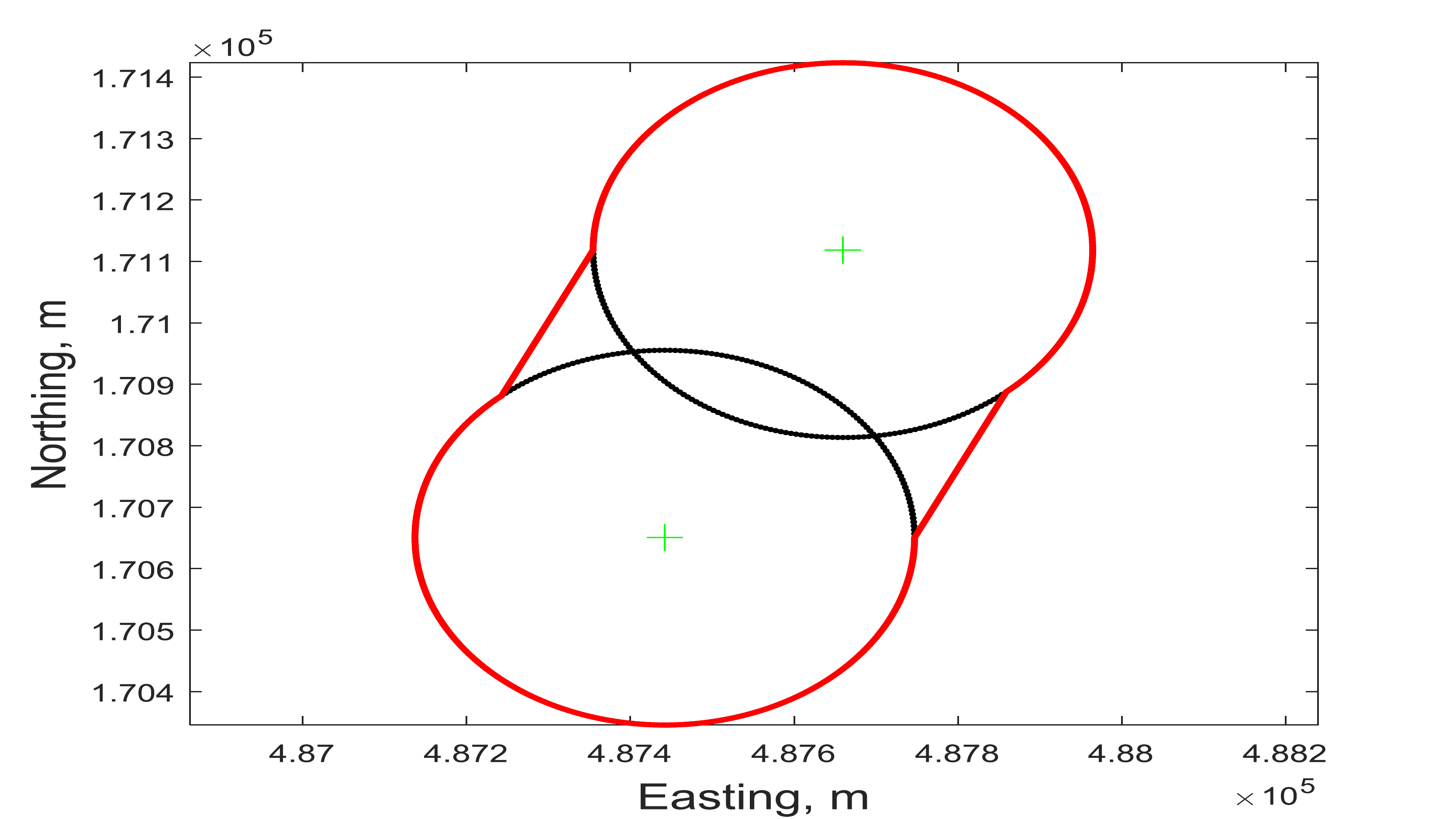}\\
\includegraphics[width=3.2in,height=2.4in]{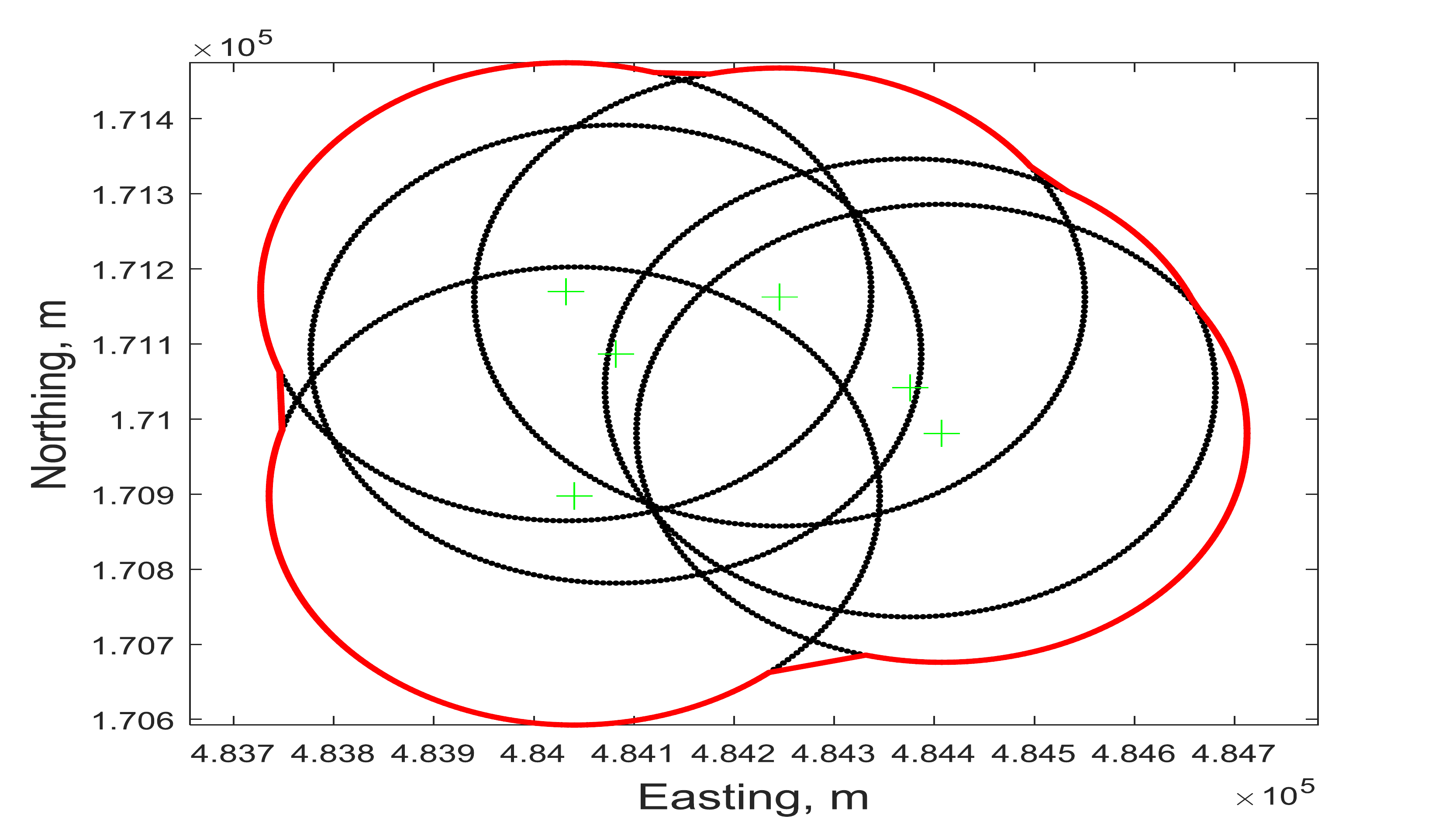}
\includegraphics[width=3.2in,height=2.4in]{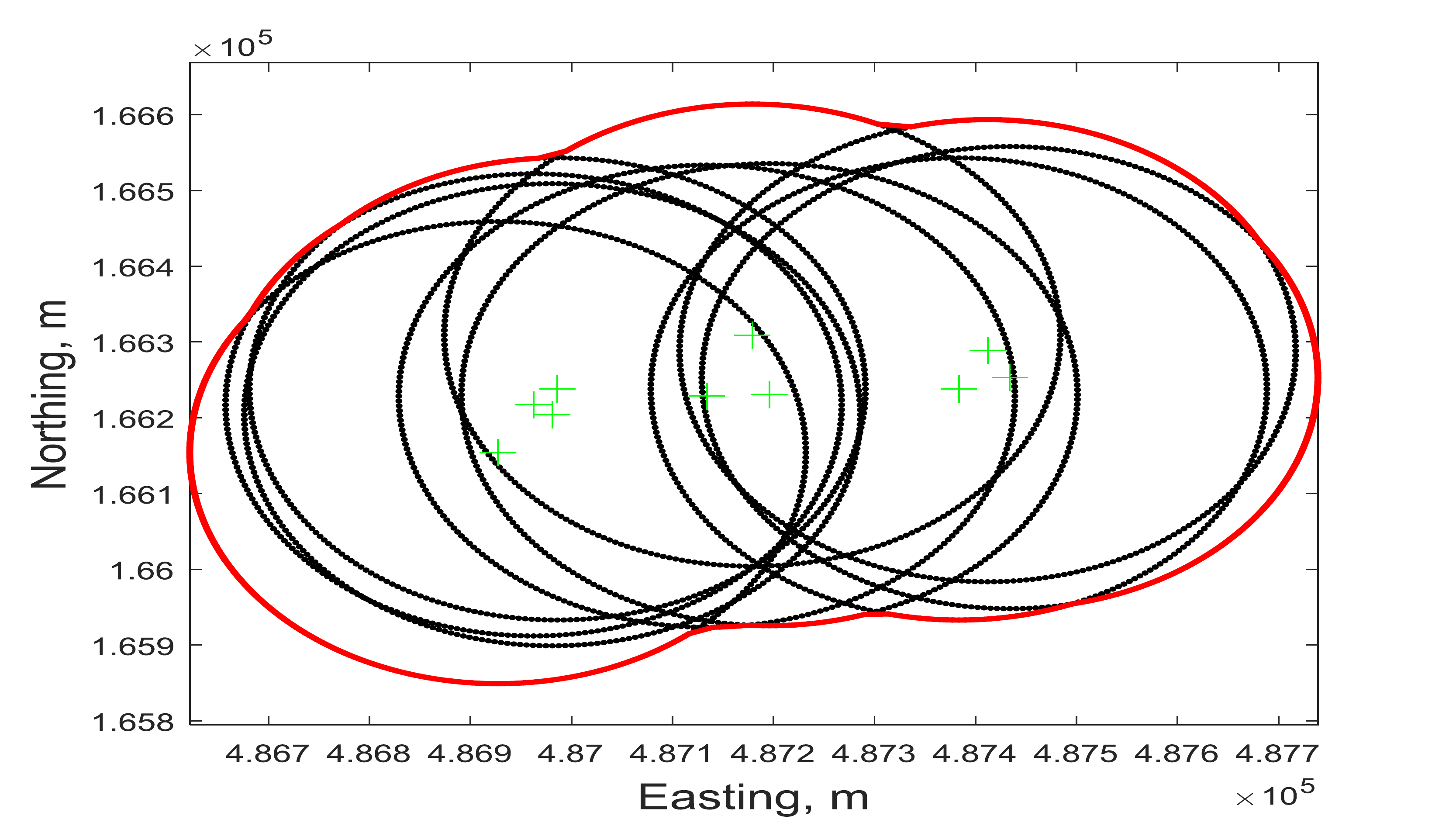}
\caption{The six distinct shapes that represent clusters of feeders, where the green crosses correspond to feeder locations and the red outline defines the shape. The black circles centred at the feeder locations have a radius of $r=305$ m. The union of these six red shapes is depicted in Figure \ref{grid}. \label{feedclus}}
\end{center}
\end{figure}

\begin{figure}
\begin{center}
\includegraphics[width=3.2in,height=2.4in]{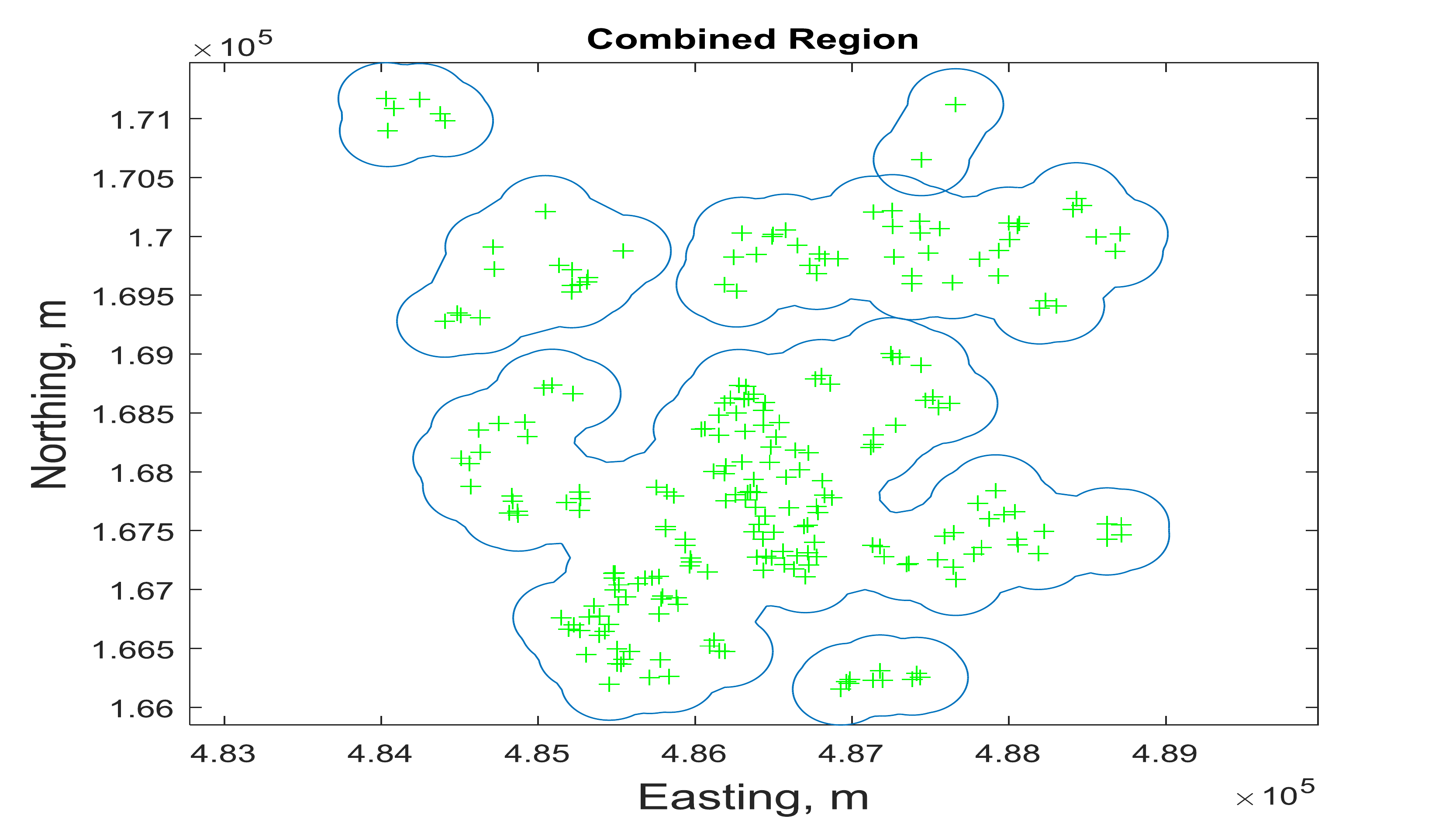}
\includegraphics[width=3.2in,height=2.4in]{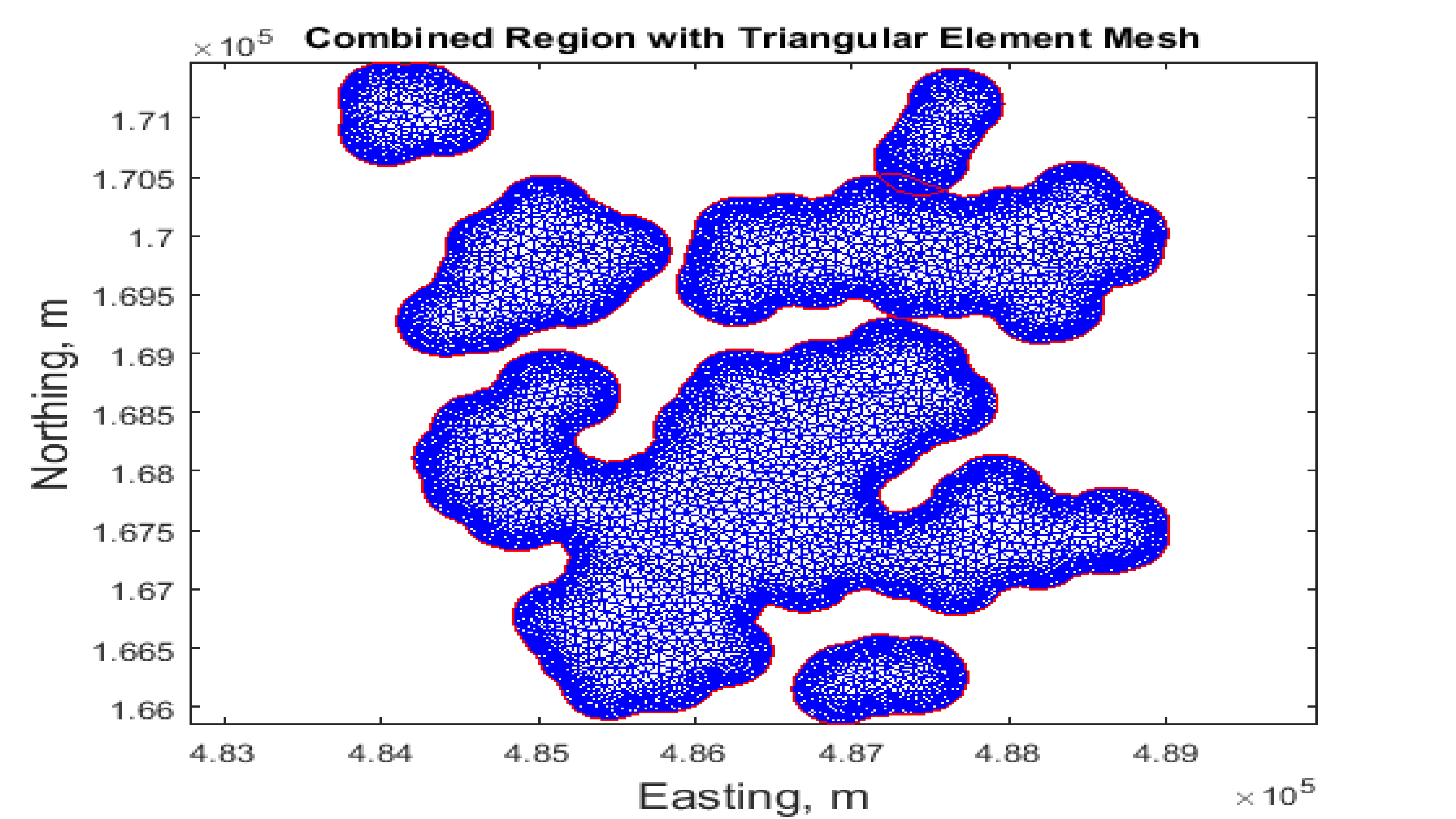}
\caption{Left: The union of the six shapes shown in Figure \ref{feedclus} with the feeder locations indicated by green crosses. Right: The union of the six shapes shown in Figure \ref{feedclus} with a triangular element mesh applied. This shaded area defines our FEM grid.
\label{grid}}
\end{center}
\end{figure}

By examining the feeder locations it is evident that they too are clustered into a number of distinct regions. So, we assign these points to separate zones. To ascertain appropriate groupings of the feeders, a circle of radius $r=305$ meters is formed around each feeder location. These circles are then used to signify the approximate area that feeders can influence or be influenced by other feeders. This area can be modified by changing $r$. Next, a region is defined by drawing a boundary around the circles that overlap. Using this method, six distinct shapes are created and when combined, they form our grid. See Figure \ref{feedclus} for depictions of the six separate regions. Then, in Figure \ref{grid}, left, the union of these shapes is illustrated, which represents the grid. As well, a triangular element mesh has been created and portrayed in Figure \ref{grid}, right. The boundaries shown are insulated except when shapes overlap, where open boundaries exist. The model (\ref{maine}) is applied to this grid using the FEM.


\begin{figure}
\begin{center}
\includegraphics[width=2.1in,height=1.6in]{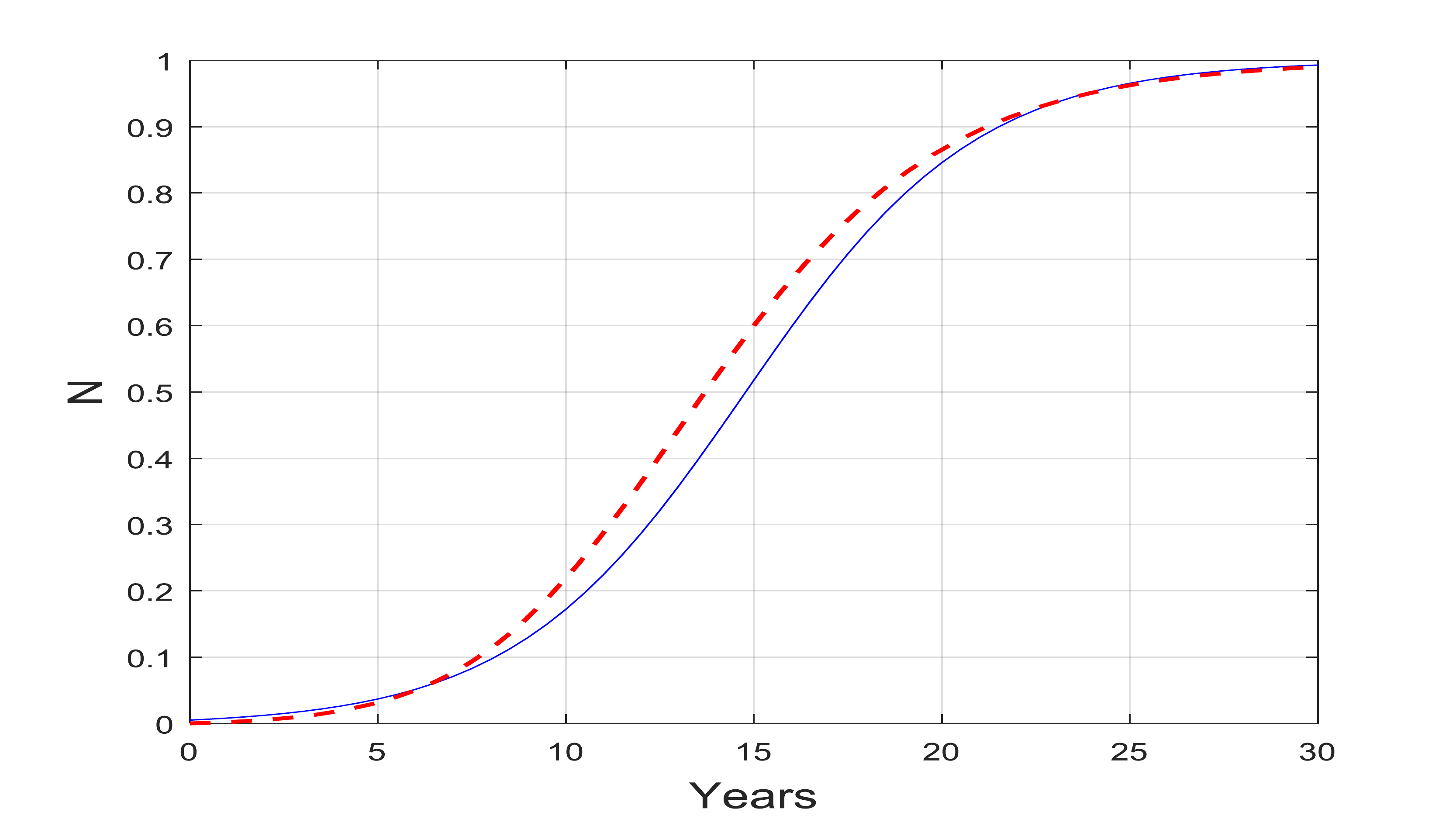}
\includegraphics[width=2.1in,height=1.6in]{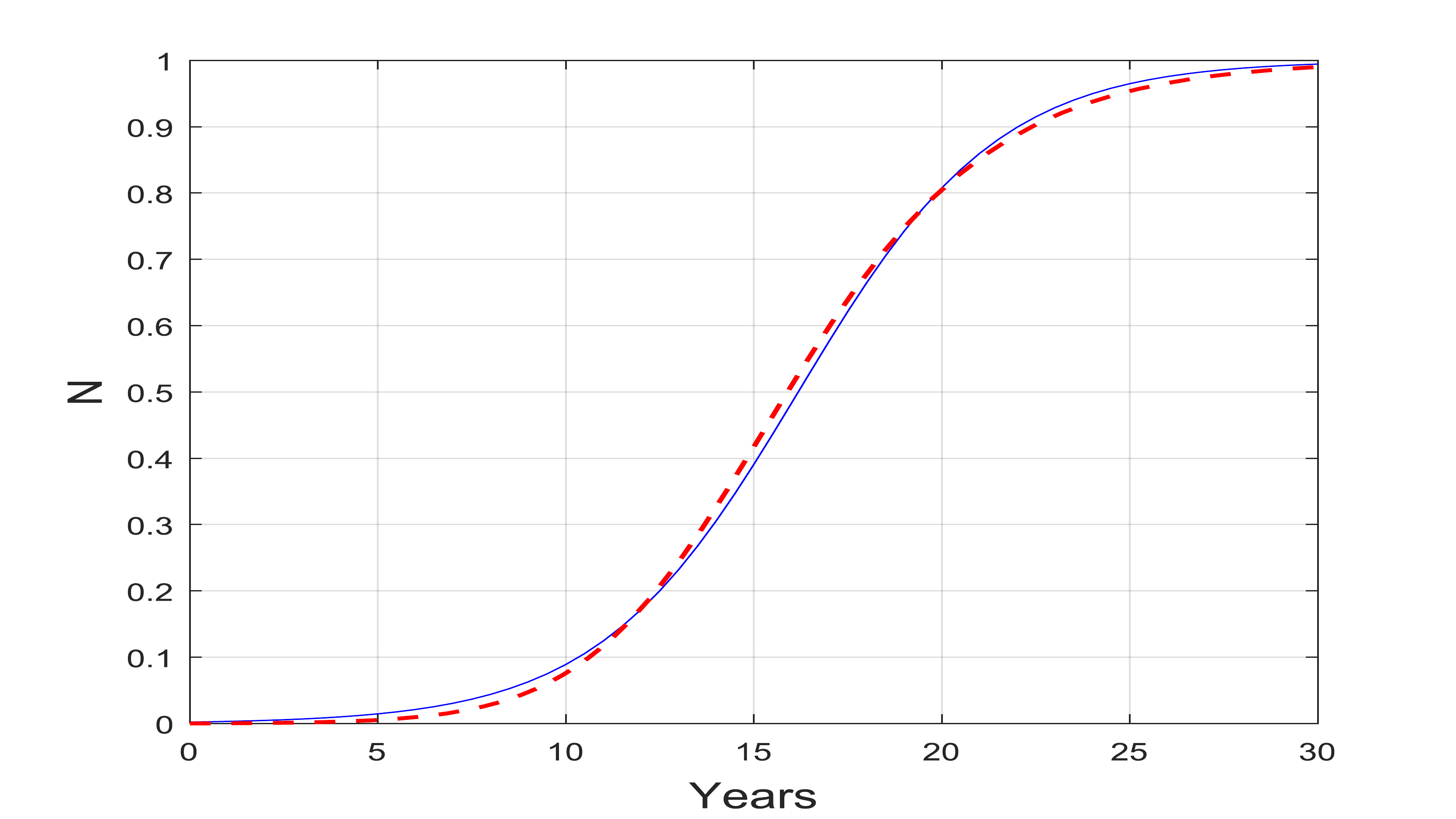}
\includegraphics[width=2.1in,height=1.6in]{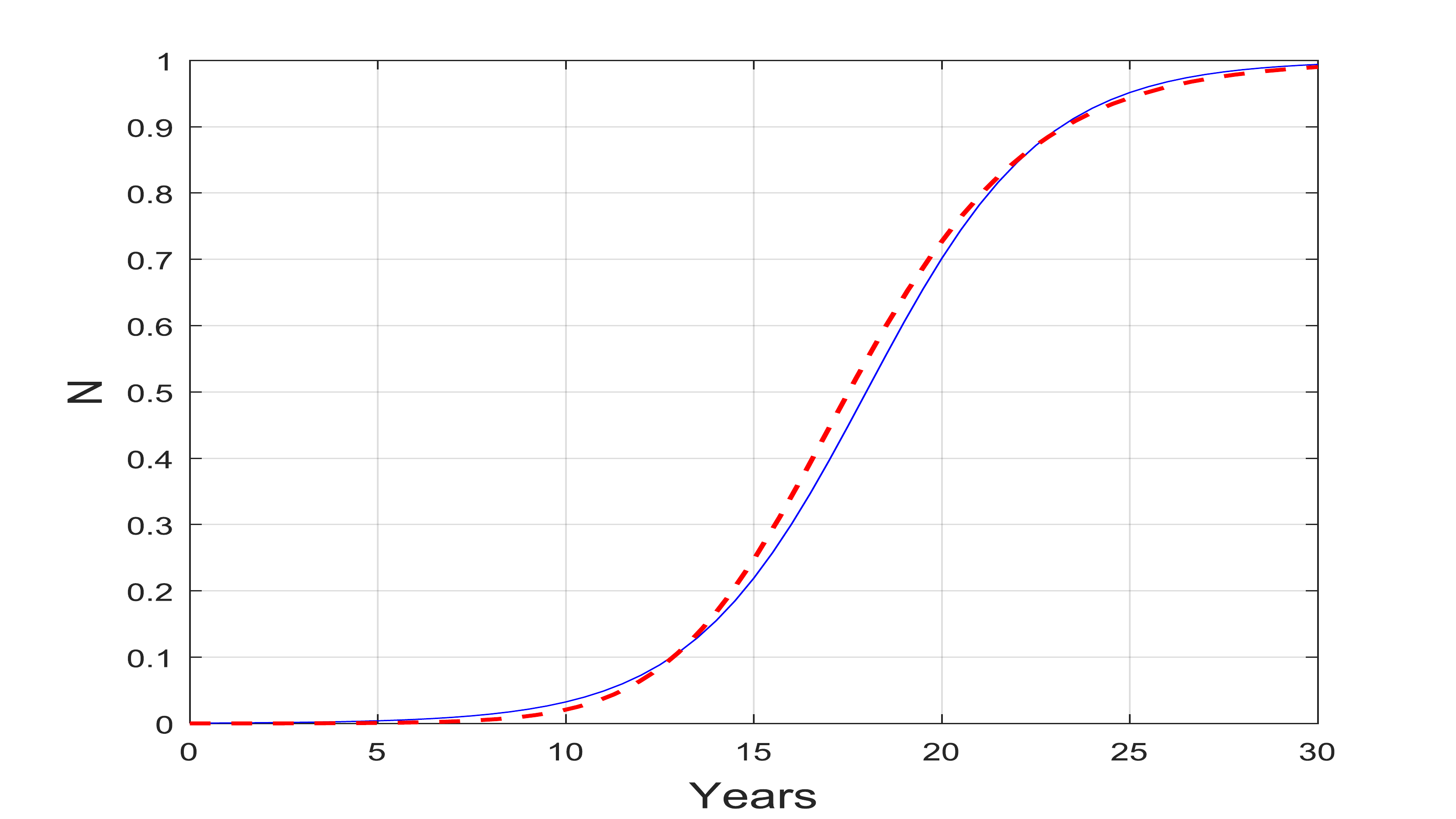}
\caption{Comparing the numerical result (blue) with the analytical solution (red dashed) when $N(t=0)=0$ everywhere, where $p=0.001$, $q=0.325$ (left), $p=0.0001$, $q=0.376$ (middle) and $p=0.00001$, $q=0.425$ (right).
\label{comp}}
\end{center}
\end{figure}

To validate the analysis detailed in Section \ref{mod}, the numerical result found using MATLAB's pde solver is compared to (\ref{travwaveN}), where the imposed initial condition is $N(t=0)=0$ everywhere (refer to Figure \ref{grid} for depictions of the solution domain). The comparisons for the three cases $(a)$, $(b)$ and $(c)$ are given in Figure \ref{comp}, where a good agreement is observed. Note that since $N=0$ everywhere at $t=0$, then $N$ remains spatially uniform for all $t>0$. Thus, the solution curves depicted in Figure  \ref{comp} are the same for all $(x,y)$.

\subsection{The Initial Condition}
The initial condition for the simulations needs to be determined. We know the location of $52$ households with PVs installed and therefore, these initial sources of innovation will be included in the model. However, this is a small number when considering there are $9484$ households in total. Therefore, an additional $100$ properties have been randomly assigned a PV for illustrative purposes. Hence, $152$ households are used as initial sources for modelling PV adoption throughout our local network.

The focus here is now the adoption of PVs within Bracknell since PV properties are known. Although, the technique presented in this paper can be used to model other LCT types, such as EVs, assuming the adoption behaviour is dependent upon internal and external influences. For PV uptake, it is expected that the imitation effect will dominate as the spatial clustering of PV properties is a known phenomenon. For example, \citet{kwa12} demonstrated that across the United States (especially in California), a correlation between zip code and a higher concentration of PV households existed. Therefore, clusters of households, like a neighbourhood, do influence one another so that spatial groupings of PV adopters form. This can be attributed to the social interactions of neighbours or the observations of neighbour behaviour. Other potential factors to consider are that neighbours generally have similar attitudes, property sizes and incomes.

\begin{figure}
\begin{center}
\includegraphics[width=2.1in,height=1.6in]{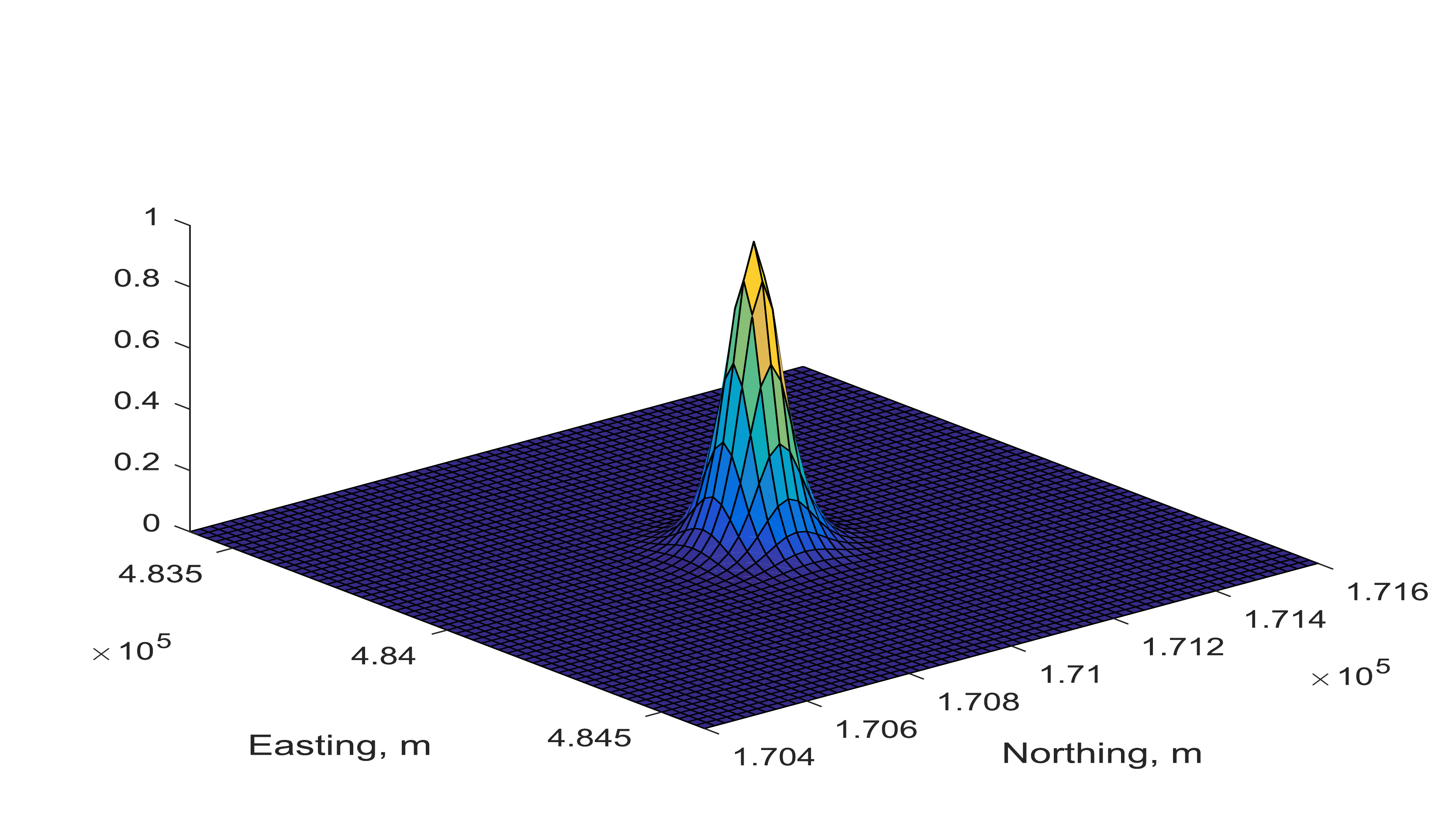}
\includegraphics[width=2.1in,height=1.6in]{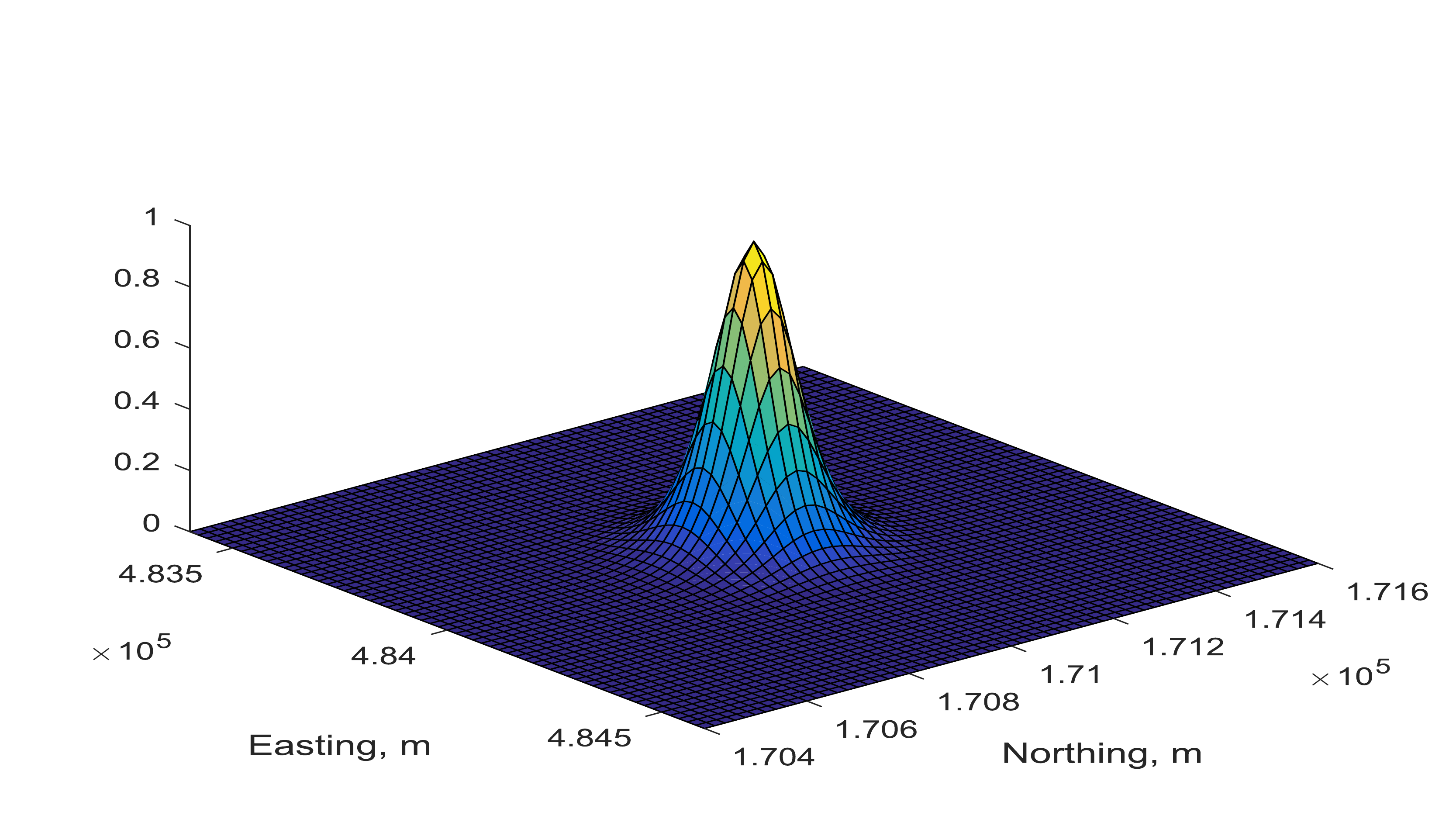}
\includegraphics[width=2.1in,height=1.6in]{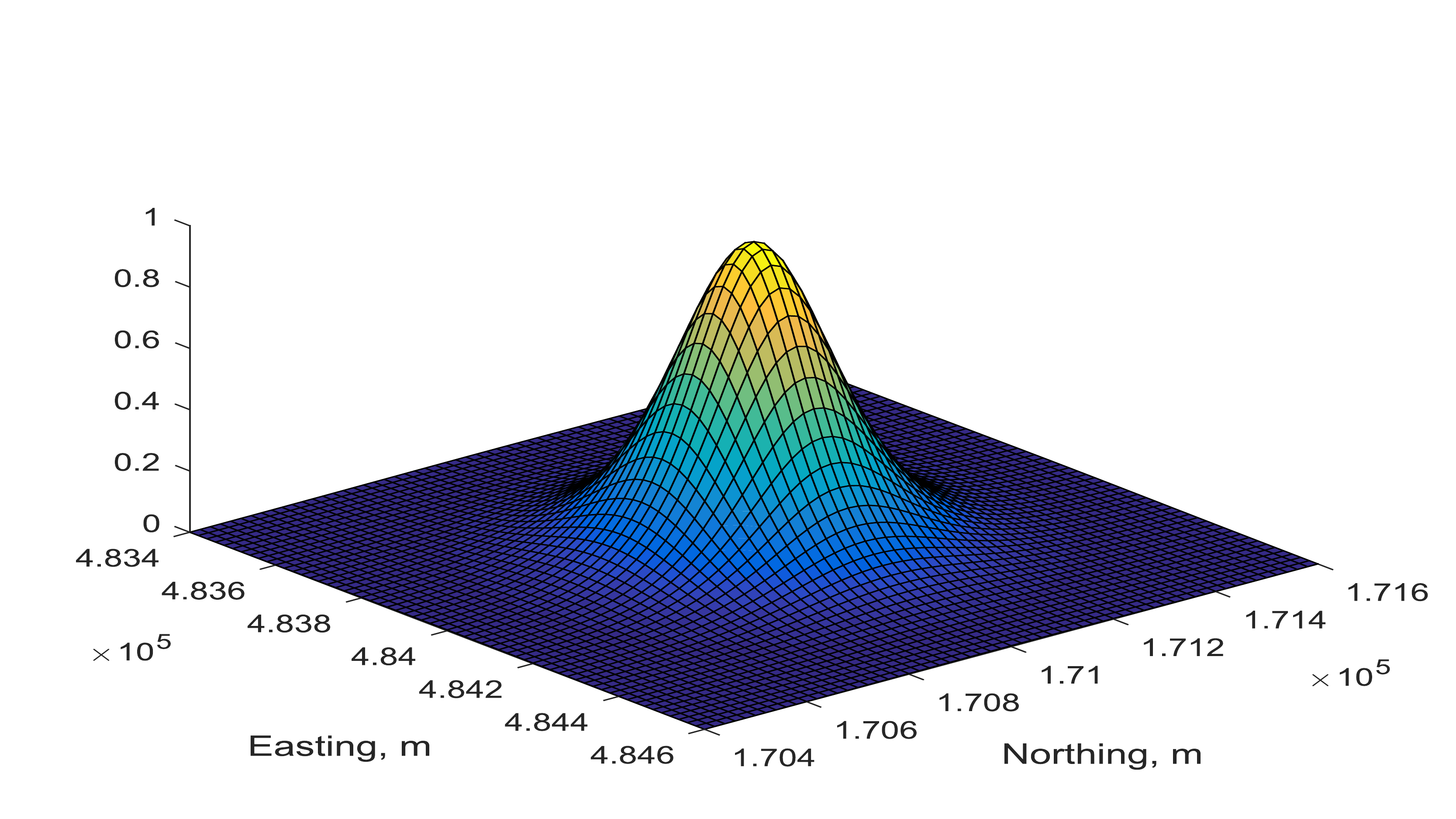}
\caption{Sources at the feeder location $(484000,171000)$ with $0-40$ households (left), $40-80$ households (middle), $80-120$ households (right). Grid size: 1200m by 1200m.
\label{init}}
\end{center}
\end{figure}

A source at the feeder location $(x_i,y_i)$ is defined by
\beq
s_i=a_i \operatorname{sech}(\beta(\sqrt{(x-x_i)^2+(y-y_i)^2}))^2,\label{source}
\eeq
where $a_i$ is the accumulative feeder adoption ratio at $t=0$ and $N(x,y,t=0)=\sum_{i=1}^{152}s_i$. Now, as $\beta$ decreases, the width of this solution increases. Therefore, we have made $\beta$ dependent upon the number of households along the feeder, choosing $\beta=0.006,0.012,0.018$ to correspond to the feeder population $0-40$ households, $40-80$ households and $80-120$ households respectively. These values of $\beta$ are applied so to reflect the approximate geographical size of the feeder. However, this is a model parameter that can be varied. See Figure \ref{init} for an example of a source located at $(484000,171000)$ with $a_i=1$ and $\beta=0.018$ (left), $\beta=0.012$ (middle) and $\beta=0.006$ (right).

\begin{figure}
\begin{center}
\includegraphics[width=3.2in,height=2.5in]{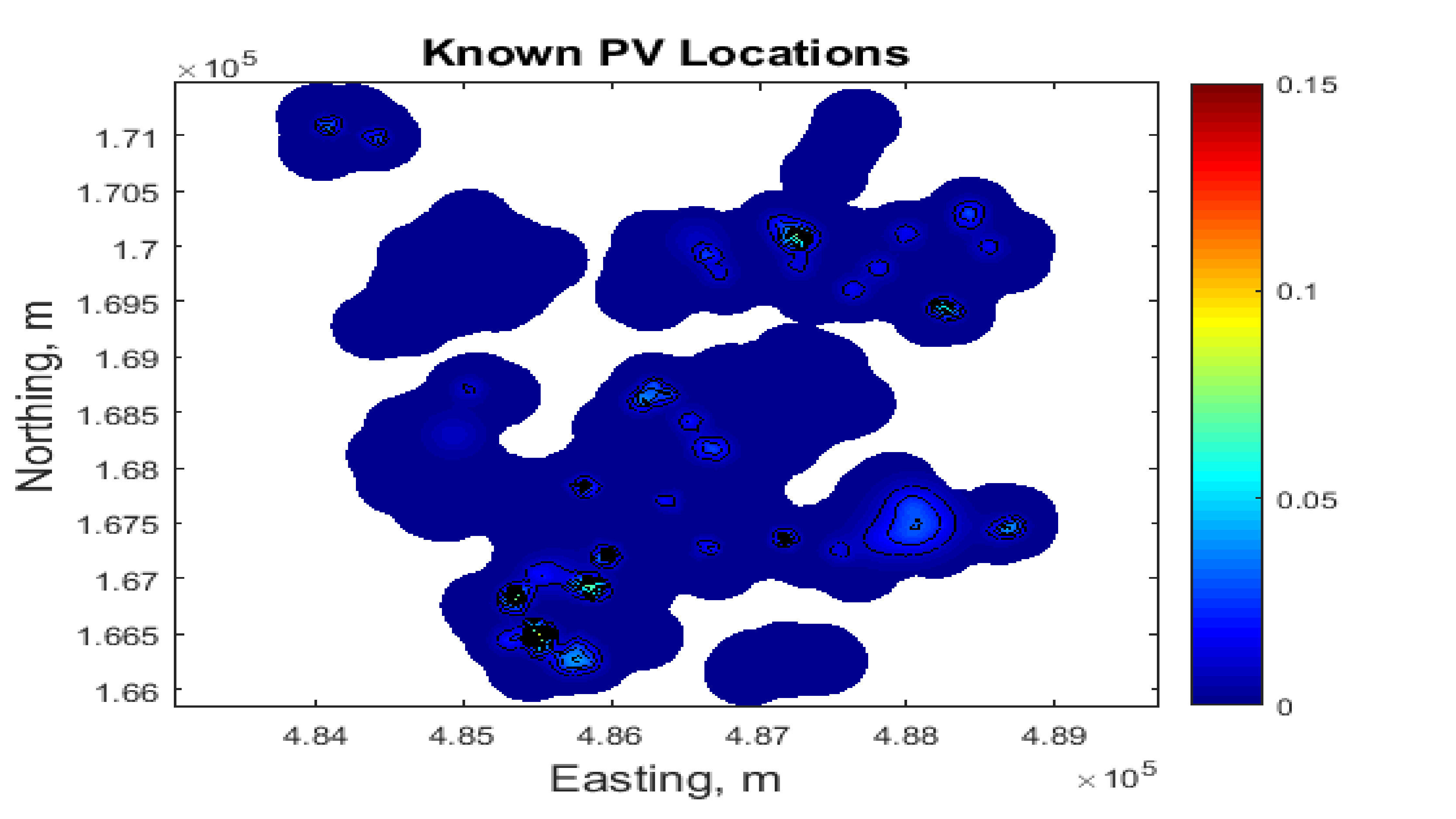}
\includegraphics[width=3.2in,height=2.5in]{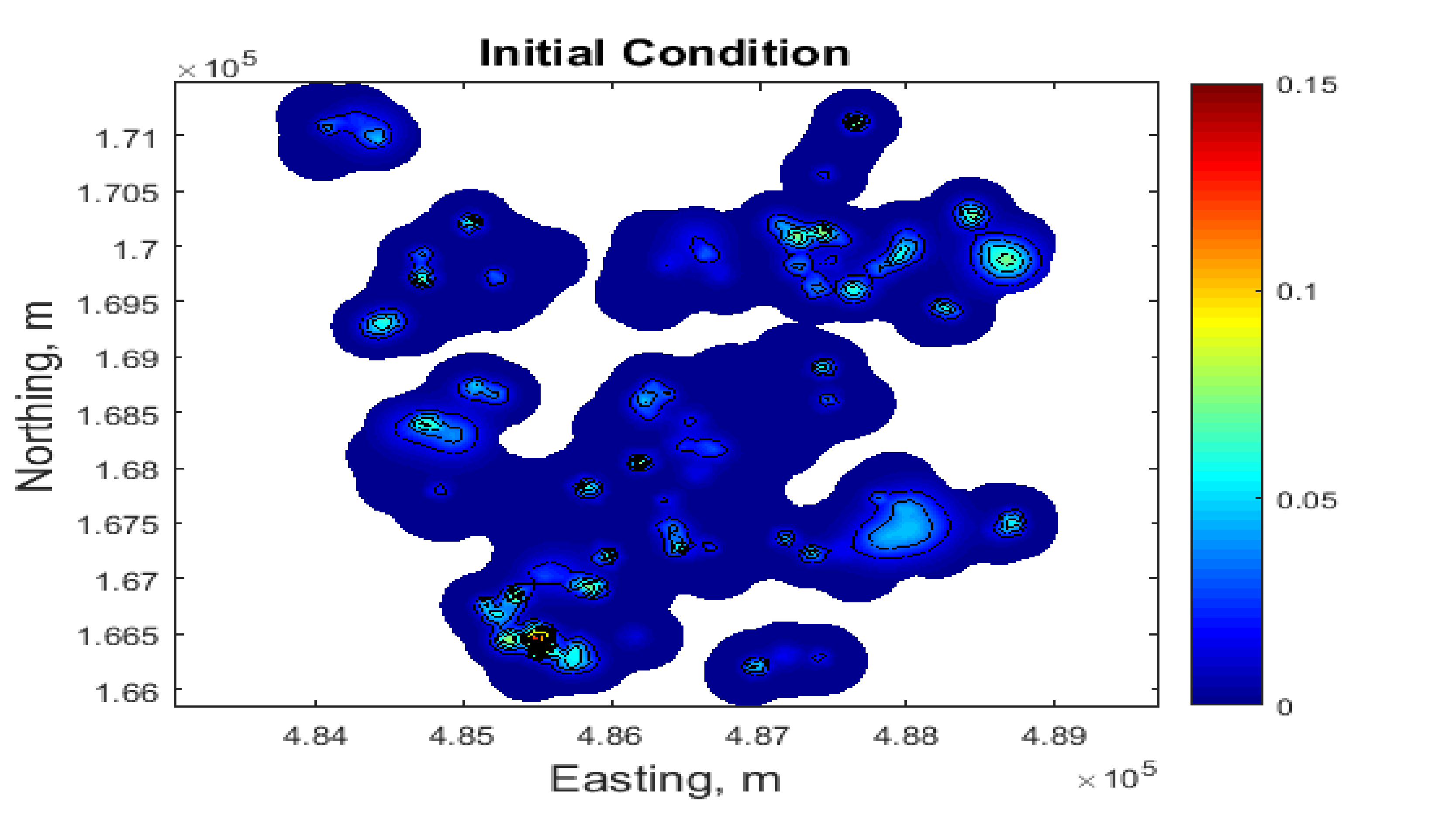}
\caption{Imposing the initial condition using (\ref{source}). Left: Known PV locations. Right: Known PV locations and the $100$ randomly assigned PV households. The legend indicates the adoption ratio, $N$.
\label{initcongrid}}
\end{center}
\end{figure}

Thus, using (\ref{source}) to define sources at the $152$ highlighted properties, the initial condition for our model is established. This solution, $N(x,y,t=0)$, is depicted in Figure \ref{initcongrid}. On the left, only the known PV locations are displayed and on the right, the combination of the known PV and the randomly assigned households is shown.

\section{Simulations}
\begin{figure}
\begin{center}
\includegraphics[width=3.2in,height=2.5in]{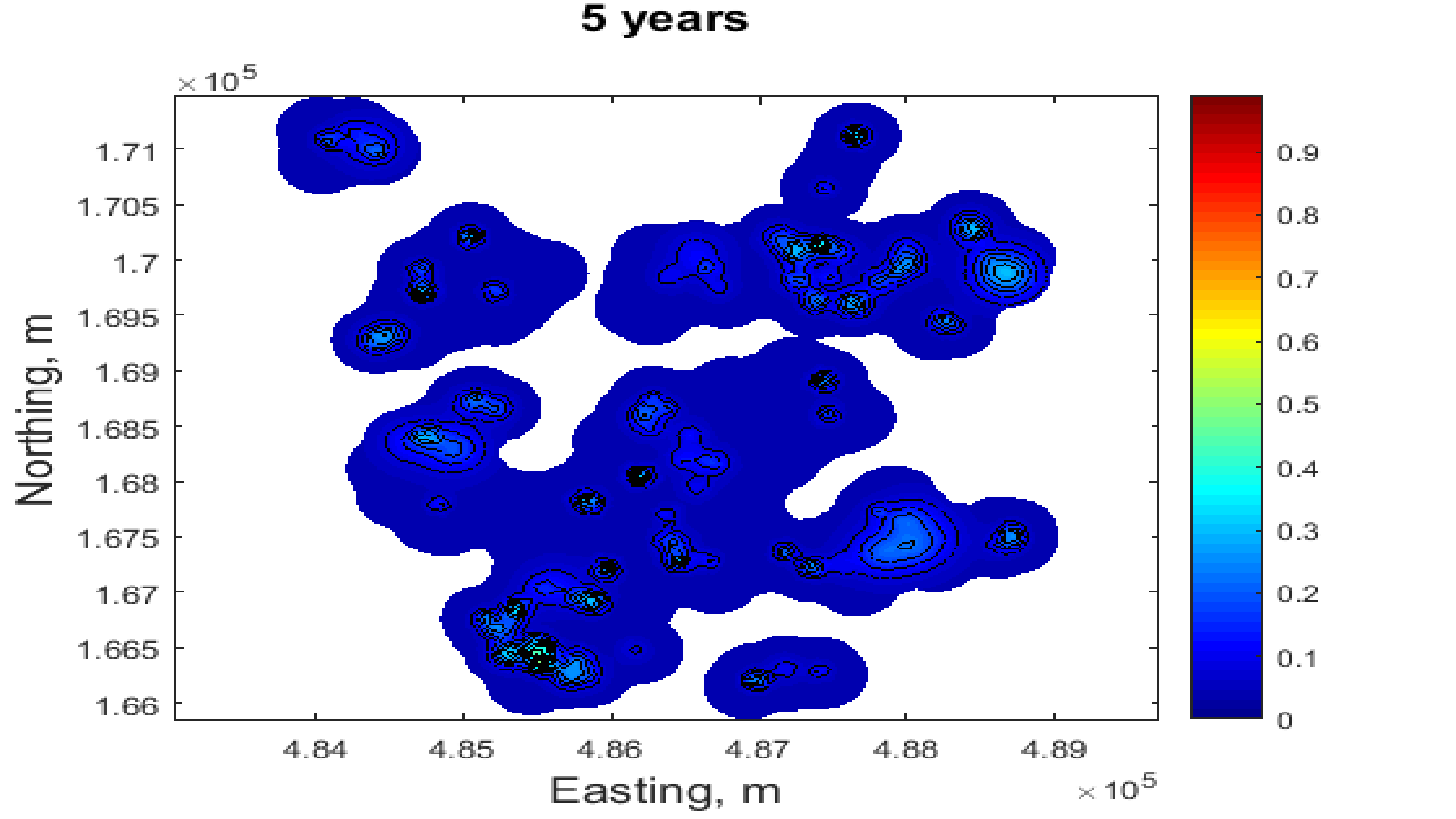}
\includegraphics[width=3.2in,height=2.5in]{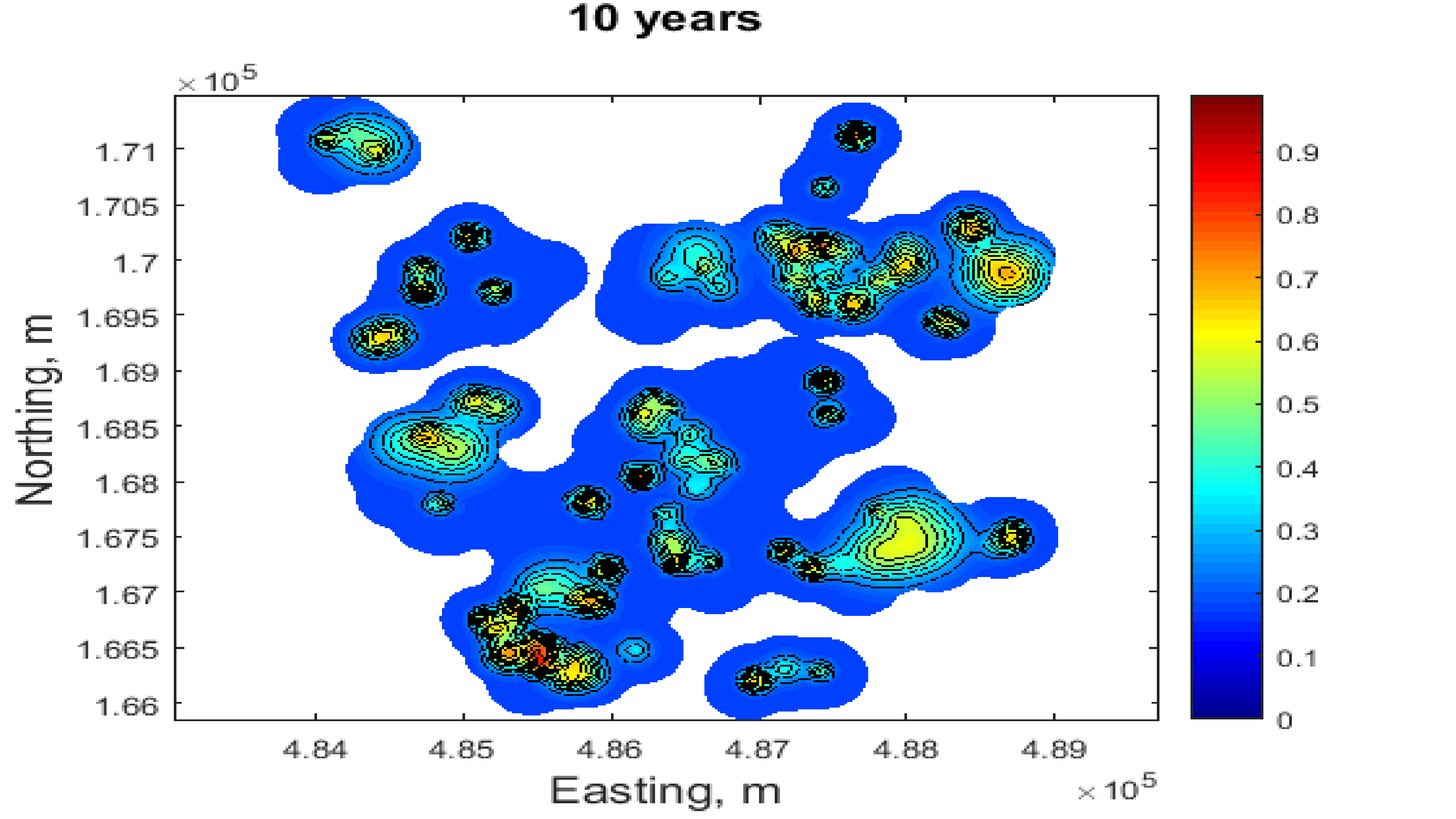}\\
\includegraphics[width=3.2in,height=2.5in]{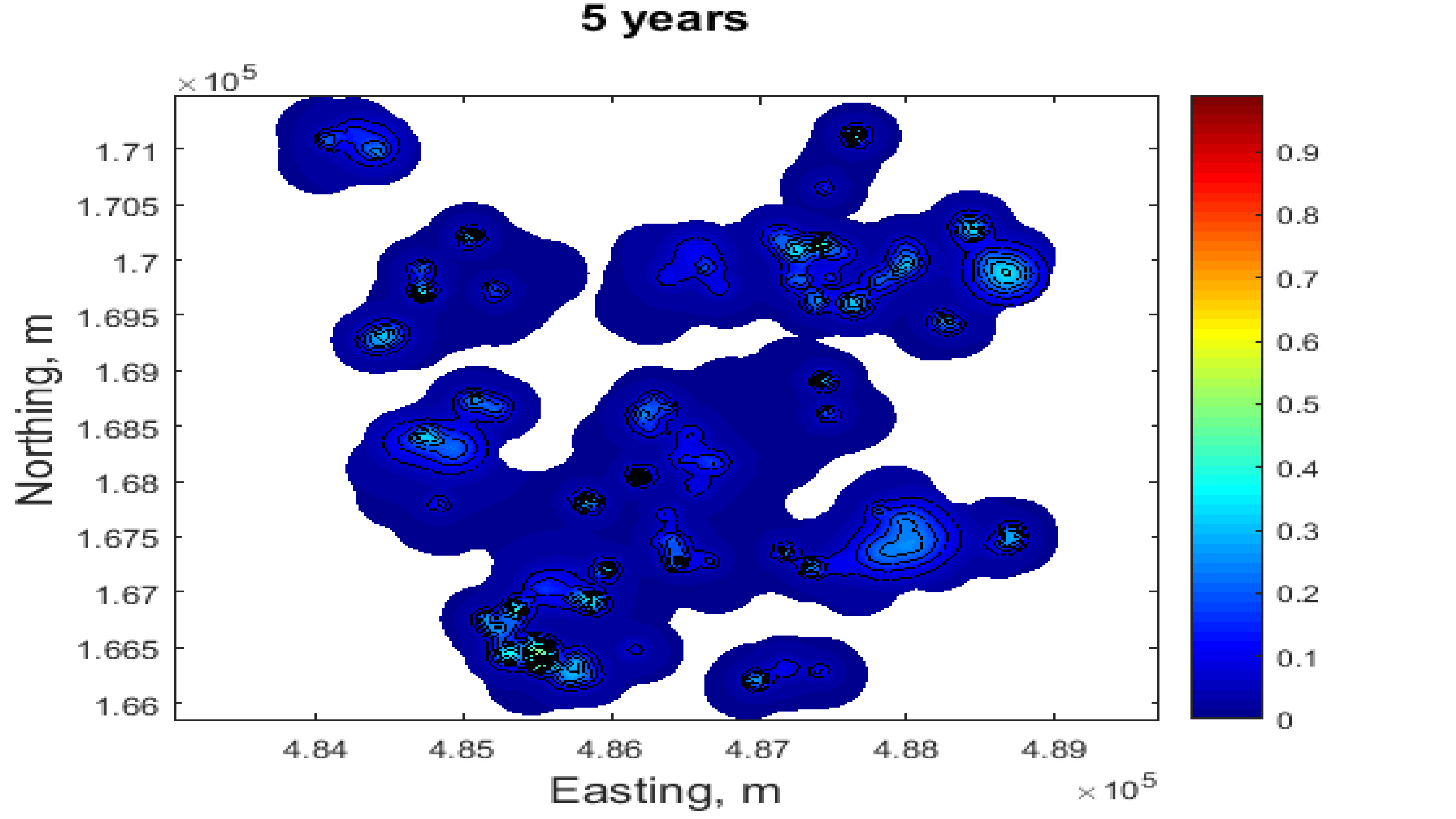}
\includegraphics[width=3.2in,height=2.5in]{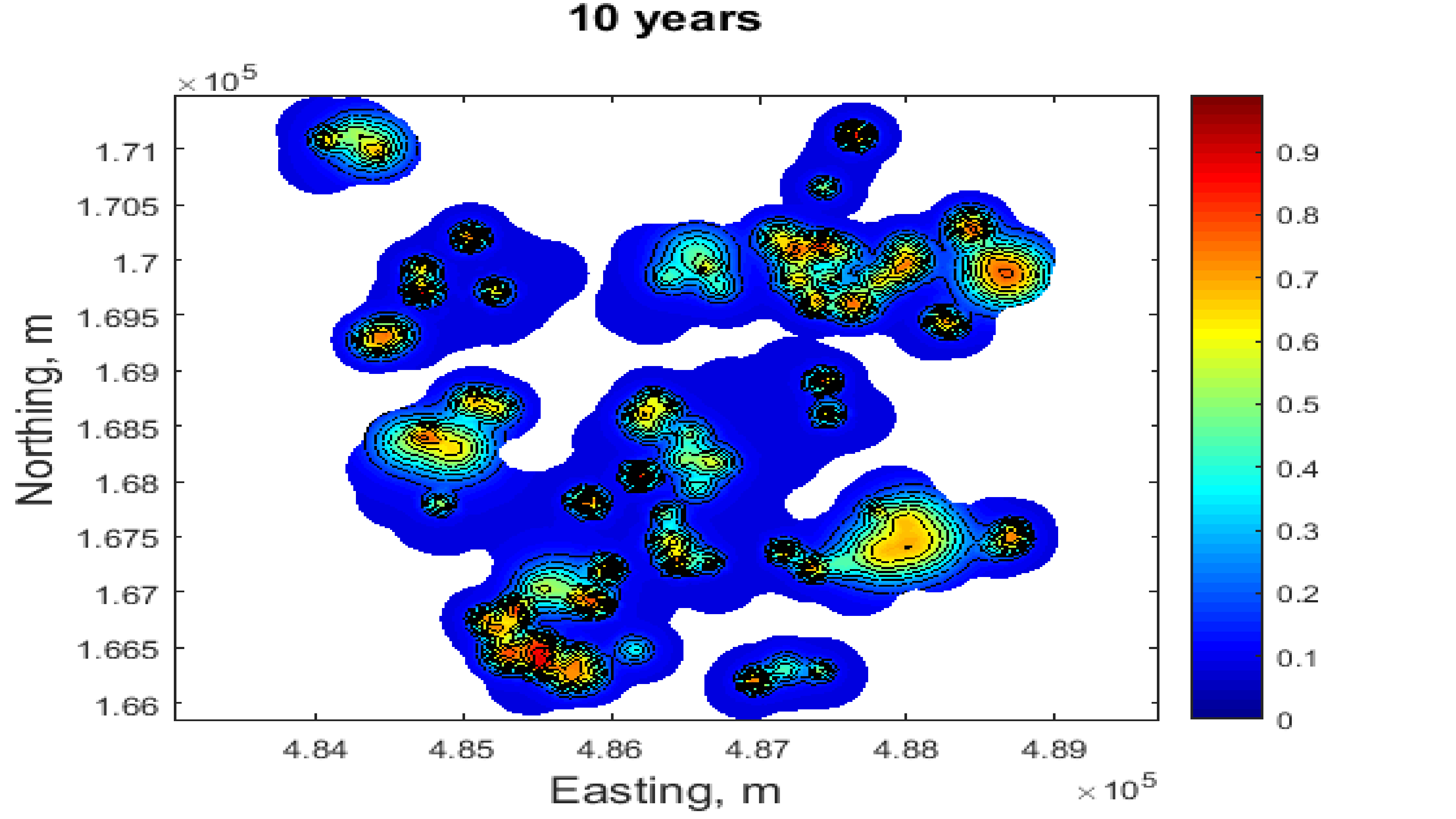}\\
\includegraphics[width=3.2in,height=2.5in]{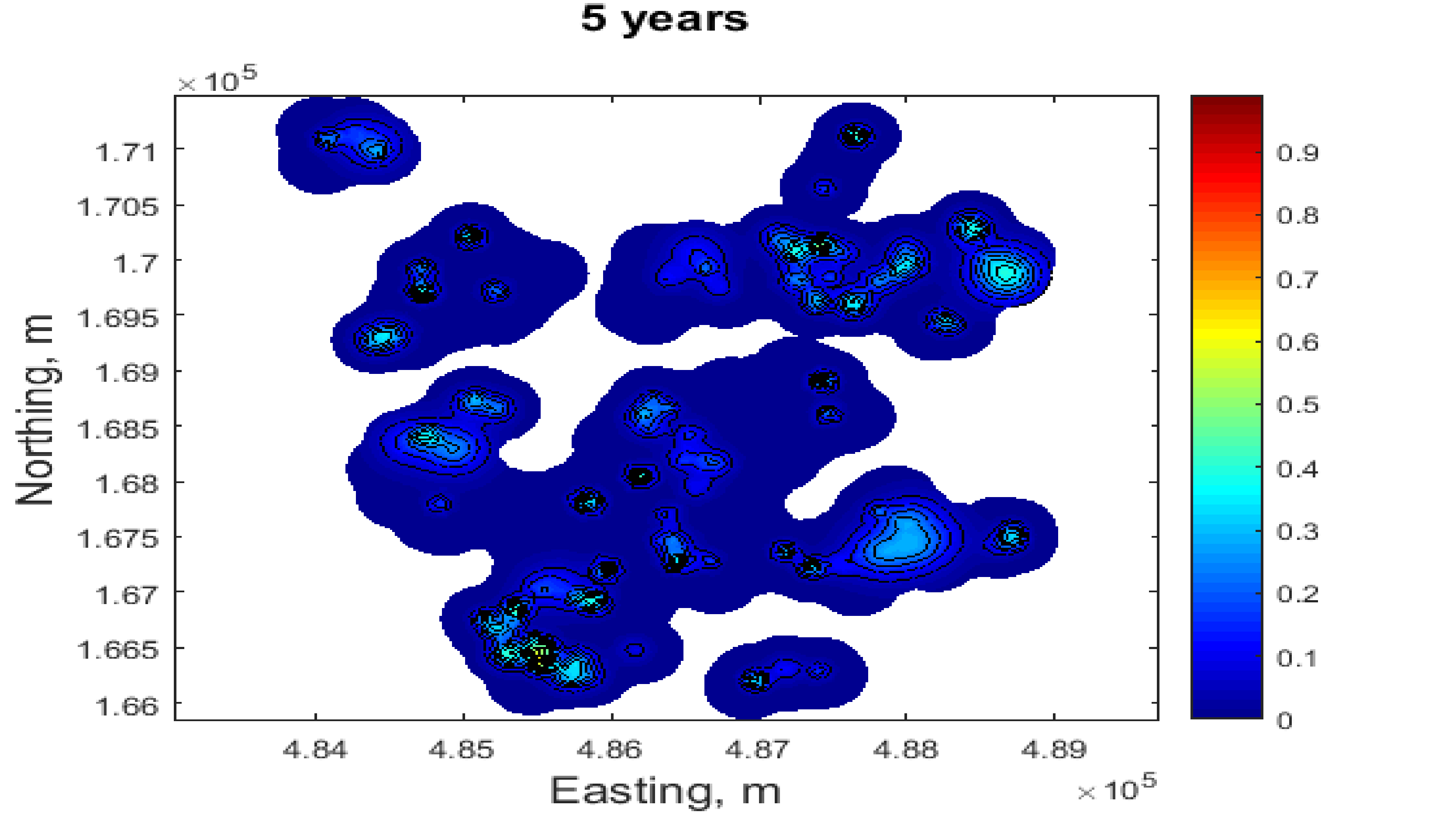}
\includegraphics[width=3.2in,height=2.5in]{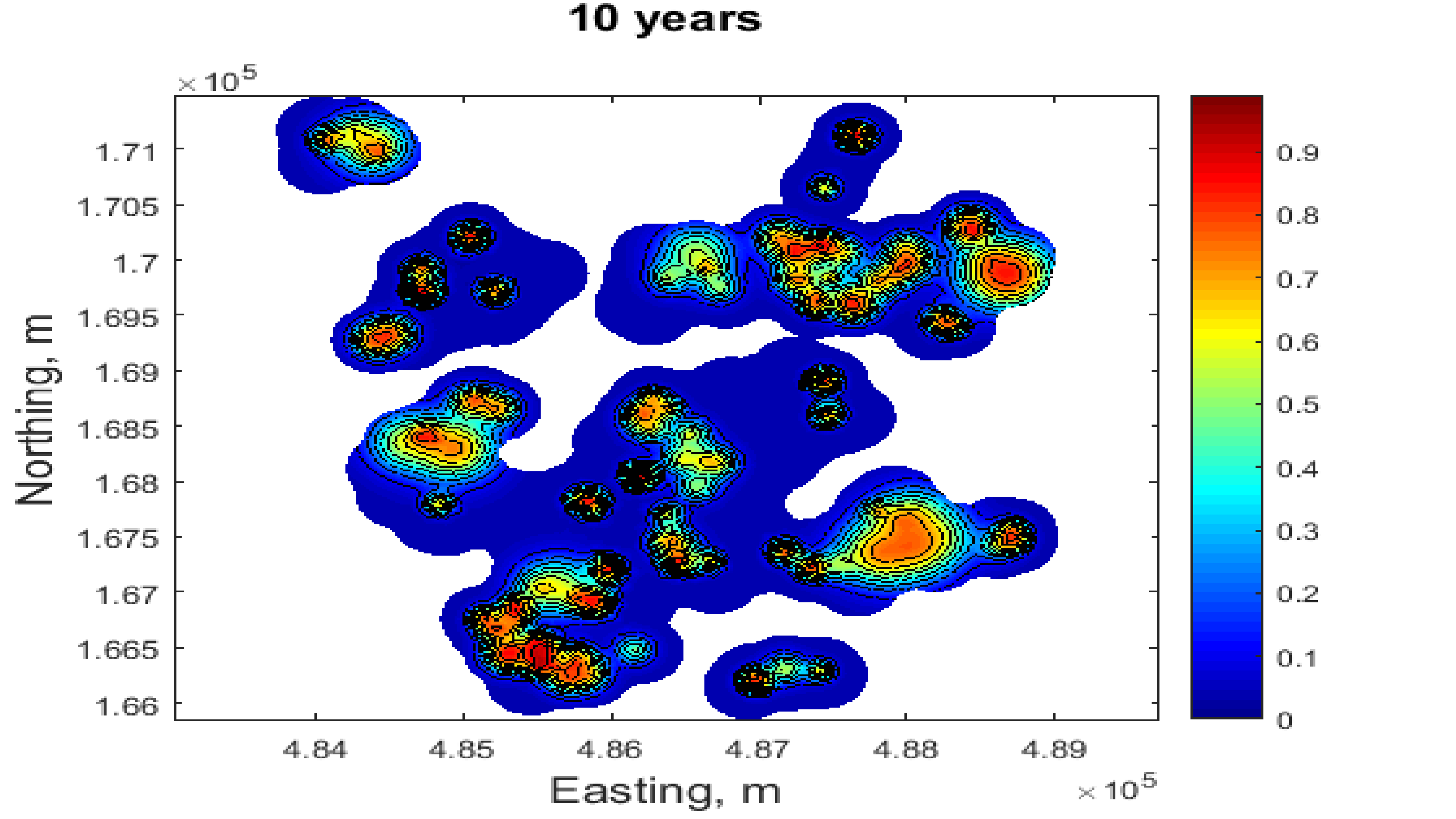}
\caption{Innovation diffusion simulations at $t=5$ years (left) and $t=10$ years (right) for case $(a)$: $p=0.001$, $q=0.325$ (top), case $(b)$: $p=0.0001$, $q=0.376$ (middle) and case $(c)$: $p=0.00001$, $q=0.425$ (bottom). The legend indicates the adoption ratio, $N$.
\label{y5y10}}
\end{center}
\end{figure}

\begin{figure}
\begin{center}
\includegraphics[width=3.2in,height=2.5in]{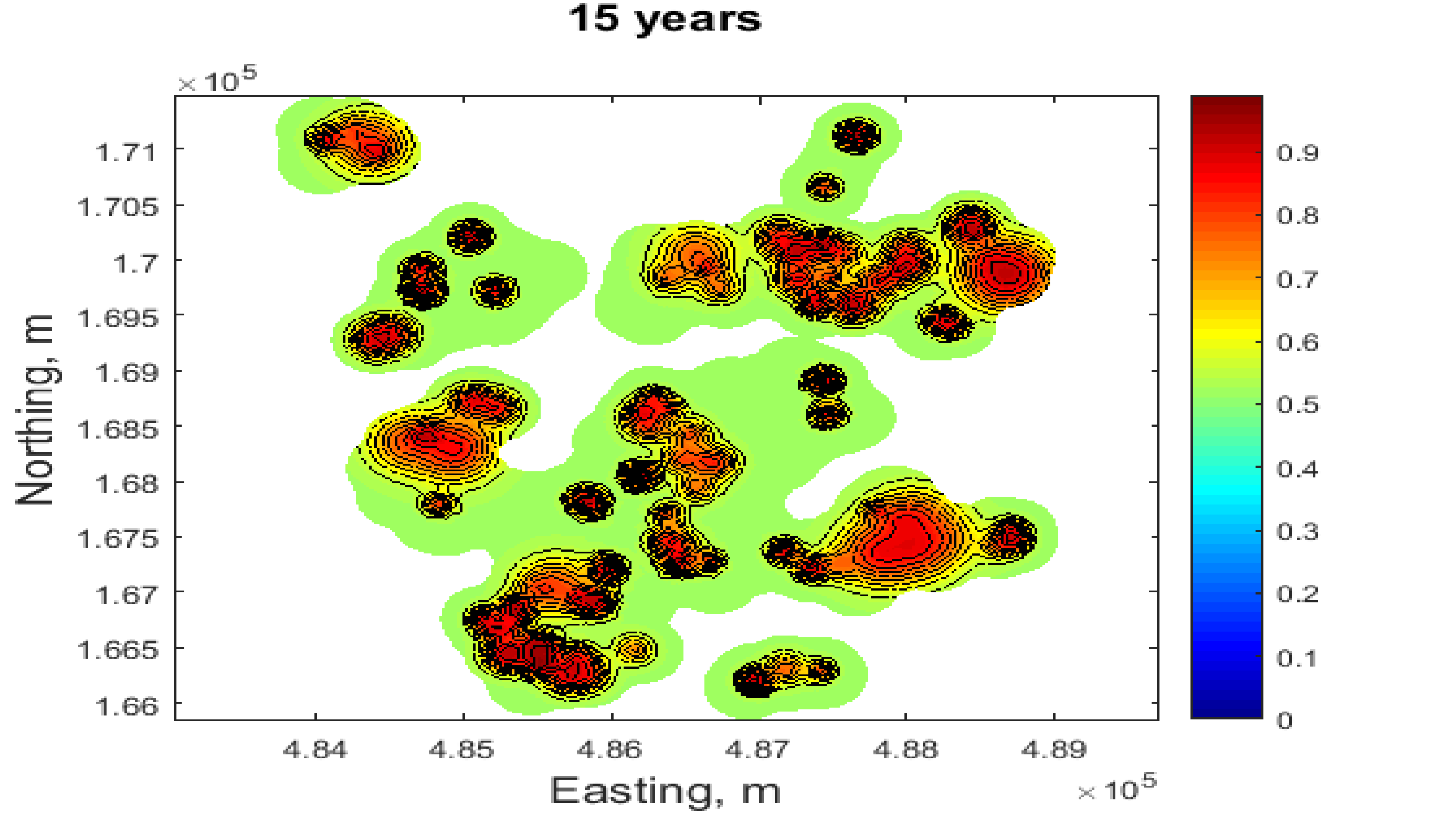}
\includegraphics[width=3.2in,height=2.5in]{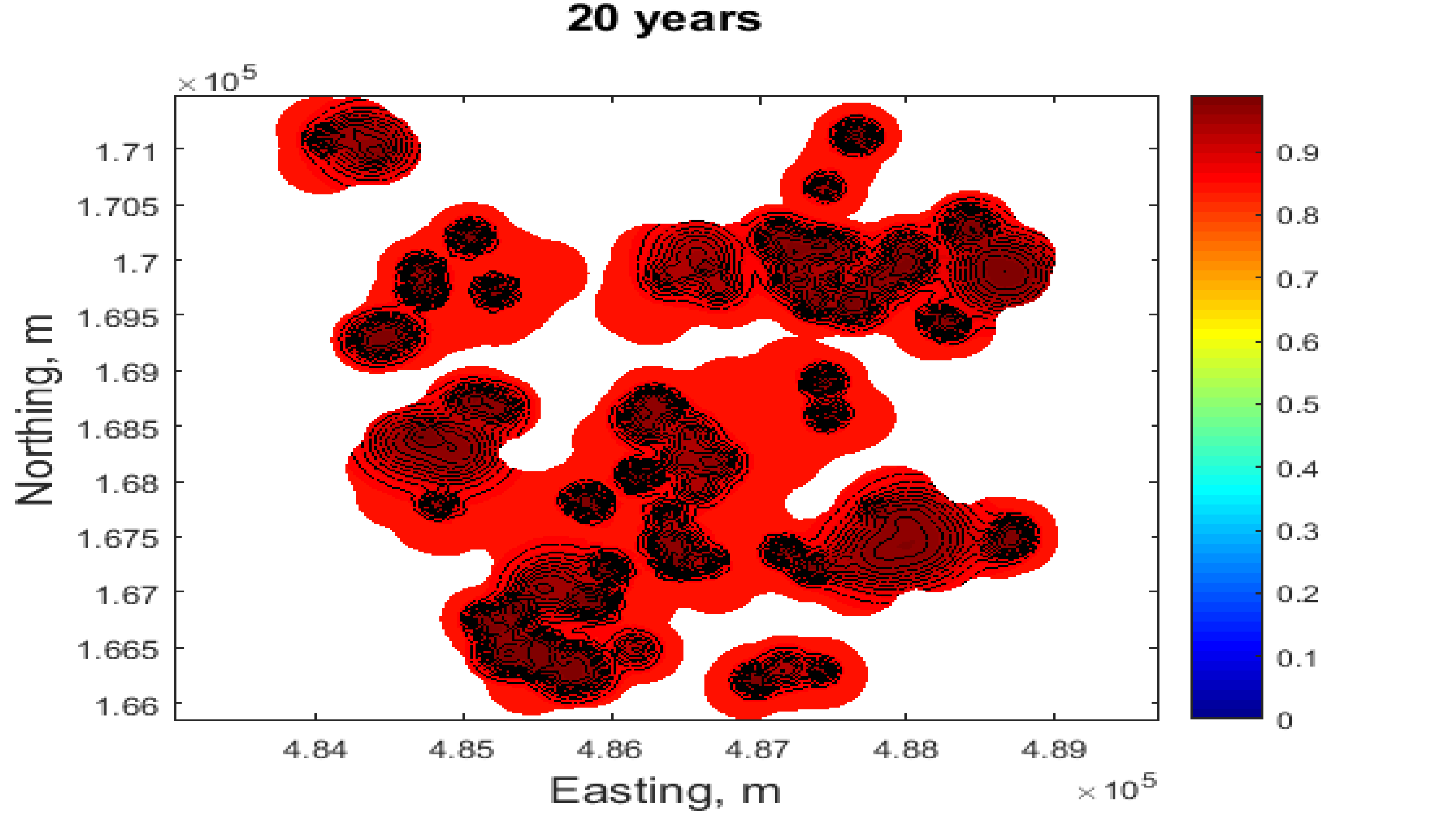}\\
\includegraphics[width=3.2in,height=2.5in]{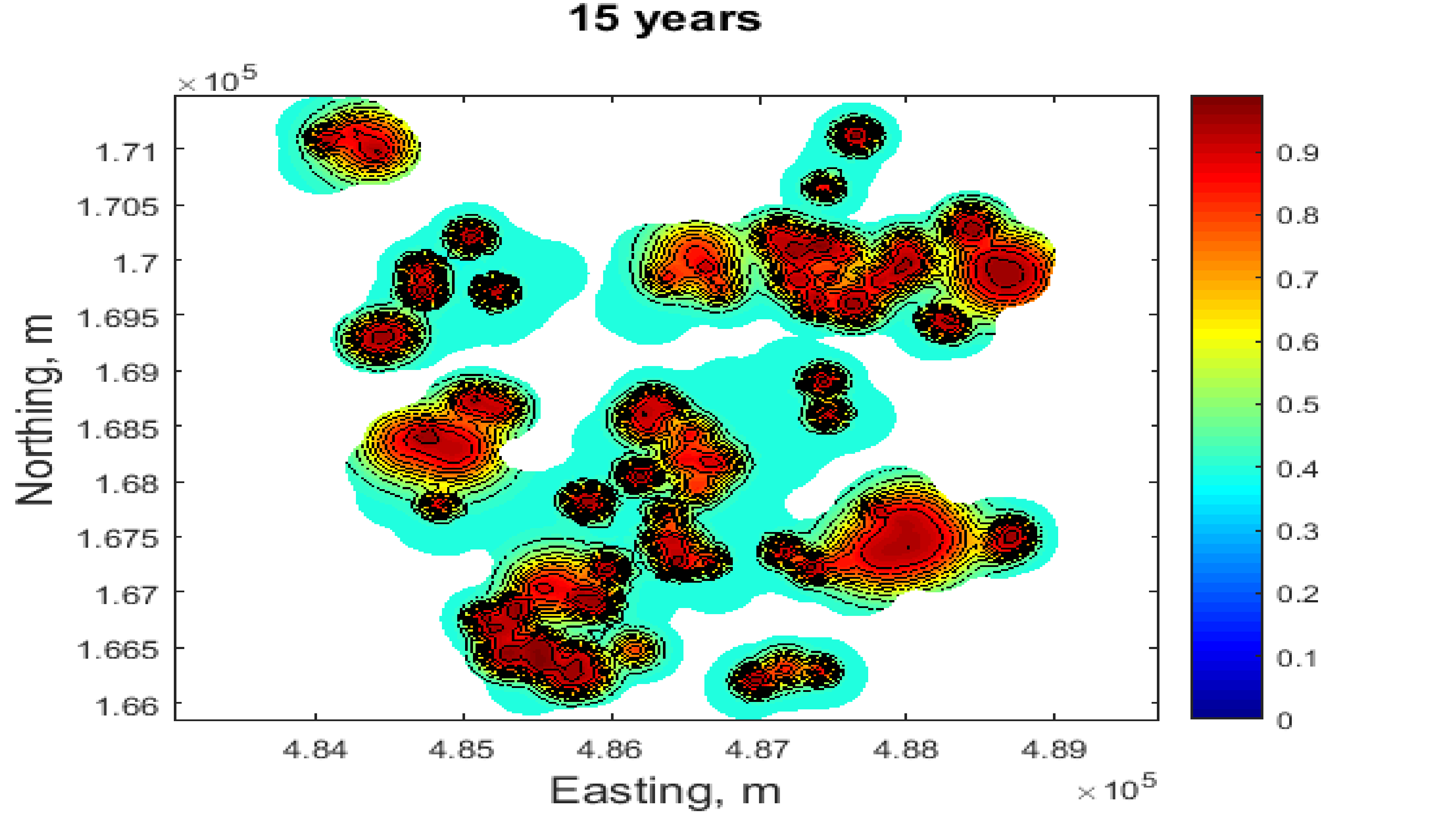}
\includegraphics[width=3.2in,height=2.5in]{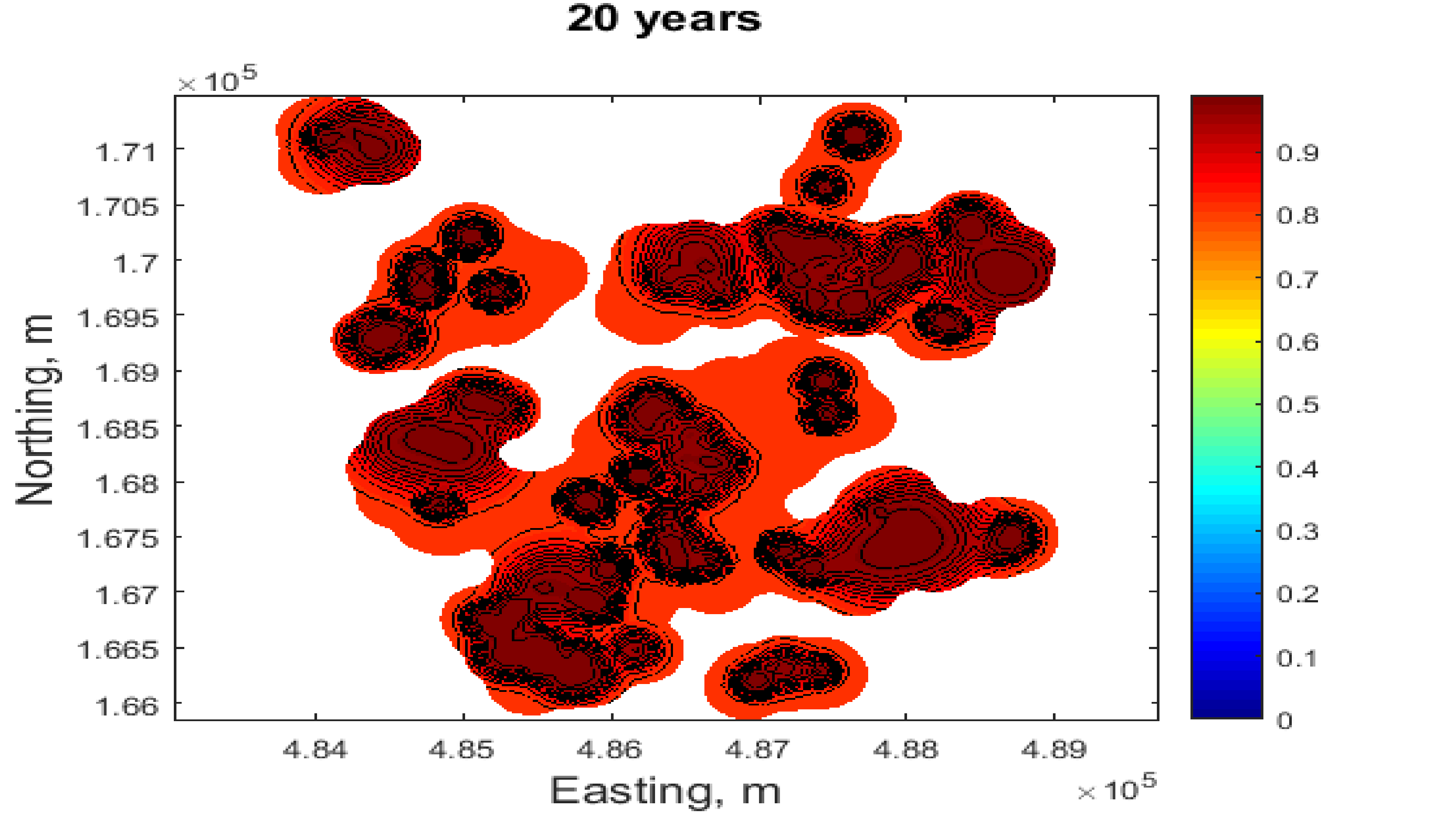}\\
\includegraphics[width=3.2in,height=2.5in]{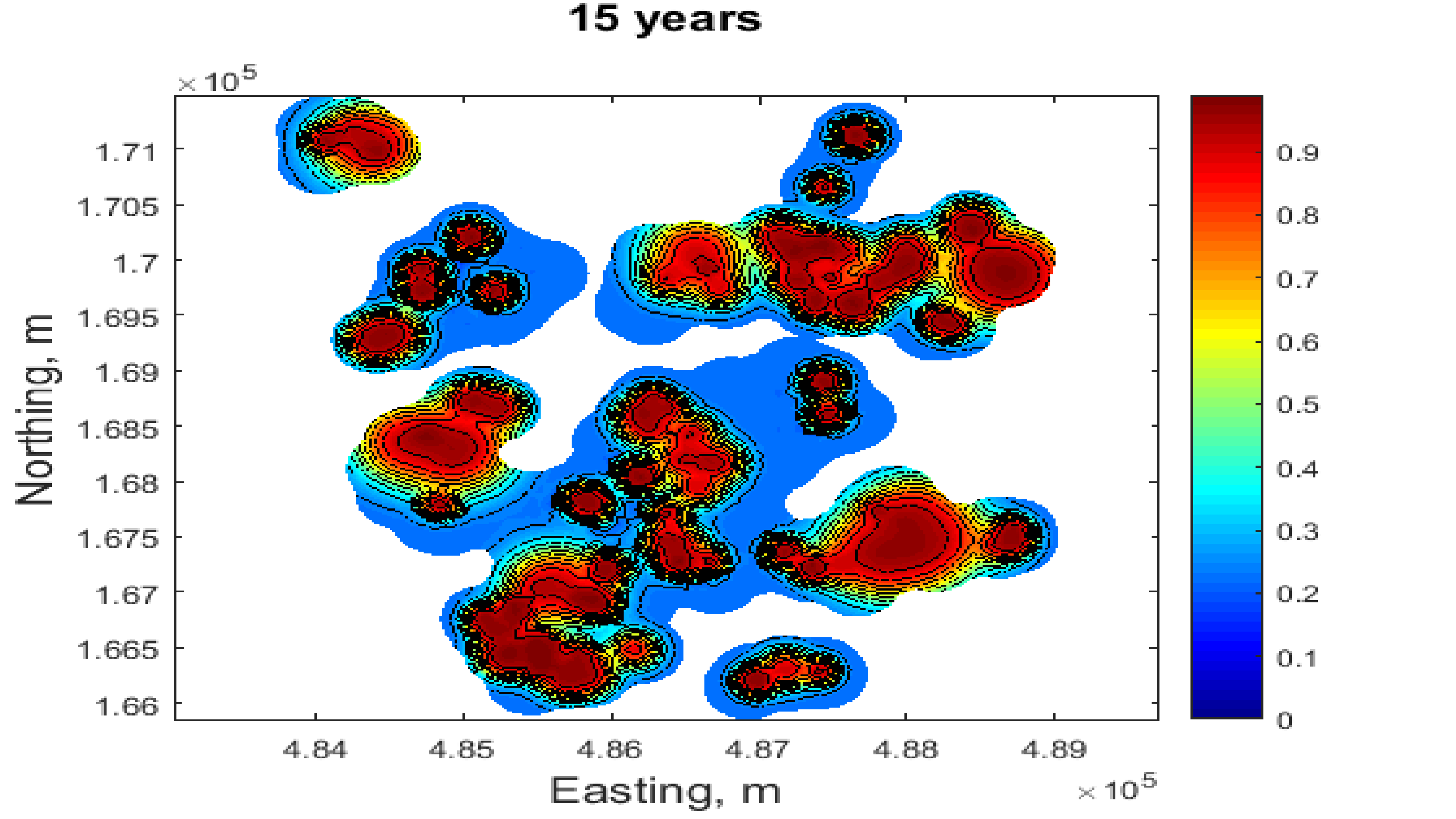}
\includegraphics[width=3.2in,height=2.5in]{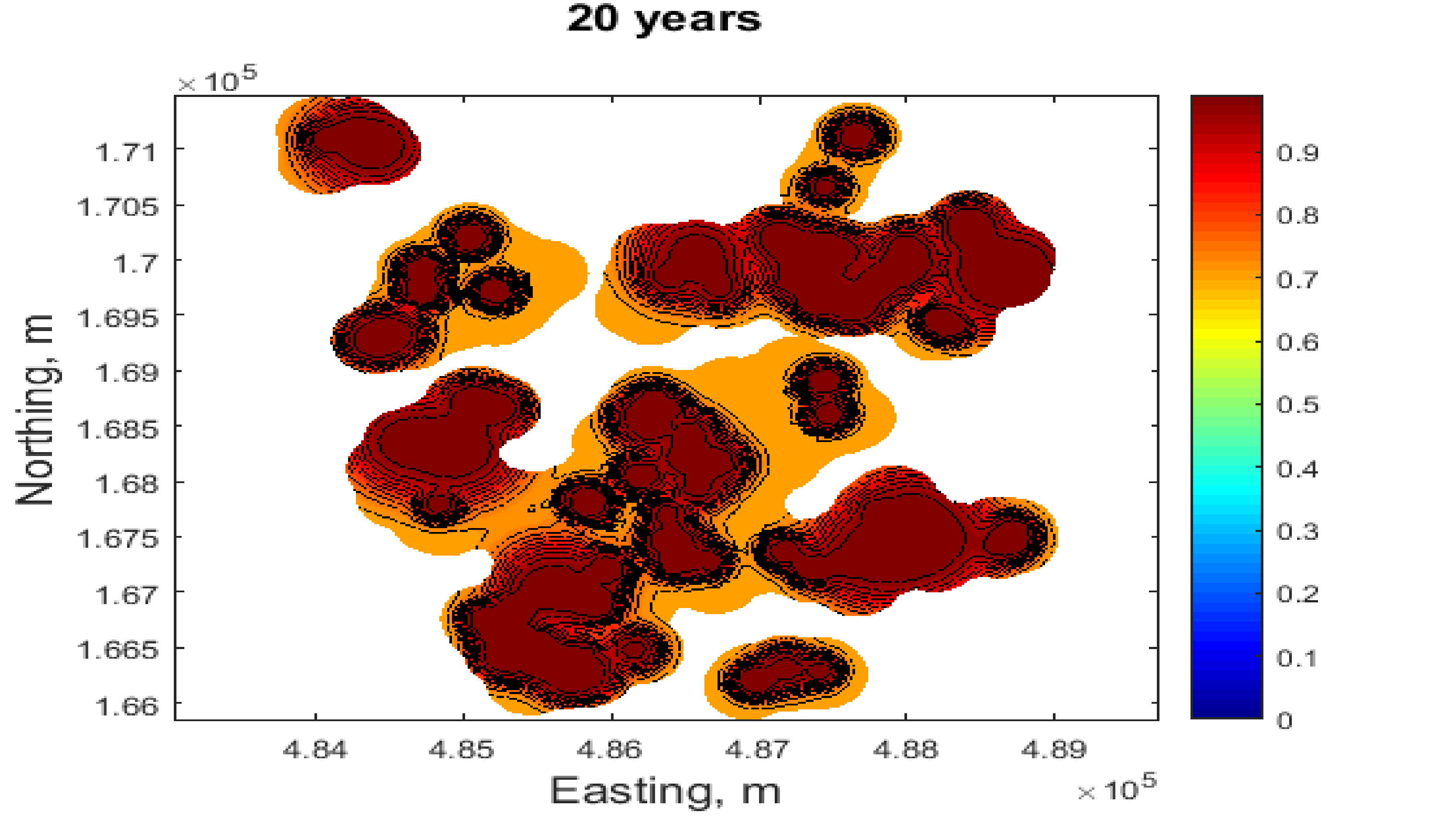}
\caption{Innovation diffusion simulations at $t=15$ years (left) and $t=20$ years (right) for $p=0.001$, $q=0.325$ (top), $p=0.0001$, $q=0.376$ (middle) and $p=0.00001$, $q=0.425$ (bottom). The legend indicates the adoption ratio, $N$.
\label{y15y20}}
\end{center}
\end{figure}

Finite element modelling has been performed with MATLAB's pde solver for the scenarios $(a)-(c)$, where the initial condition depicted in Figure \ref{initcongrid}, right was applied. In Figures \ref{y5y10} and \ref{y15y20}, the results at $t=5,10,15,20$ years are displayed. As expected, when the imitation coefficient is much larger than the innovation effect, the adoption clusters surrounding the feeder locations become notably more intense with increasing time. For instance, comparing the maps at $t=10$ years for cases $(a)$-$(c)$, it is apparent that case $(c)$ exhibits much greater adoption numbers in the areas around the feeders with sources. As $t$ becomes larger, these clusters expand and amplify. However, it should be noted that the background, overall adoption ratio (this follows the trends given in Figure \ref{comp}) is evidently greater, in particular at $t=15,20$ years, when the innovation influence is increased. Consequently, the contrast between the adoption clusters and the background level is far more apparent for case $(c)$. These observations are expected since $q$ will have a localised effect (household adoptions resulting from neighbour activity), whereas $p$ has an overall impact (external factors encouraging adoptions).

\begin{figure}
\begin{center}
\includegraphics[width=3.2in,height=2.5in]{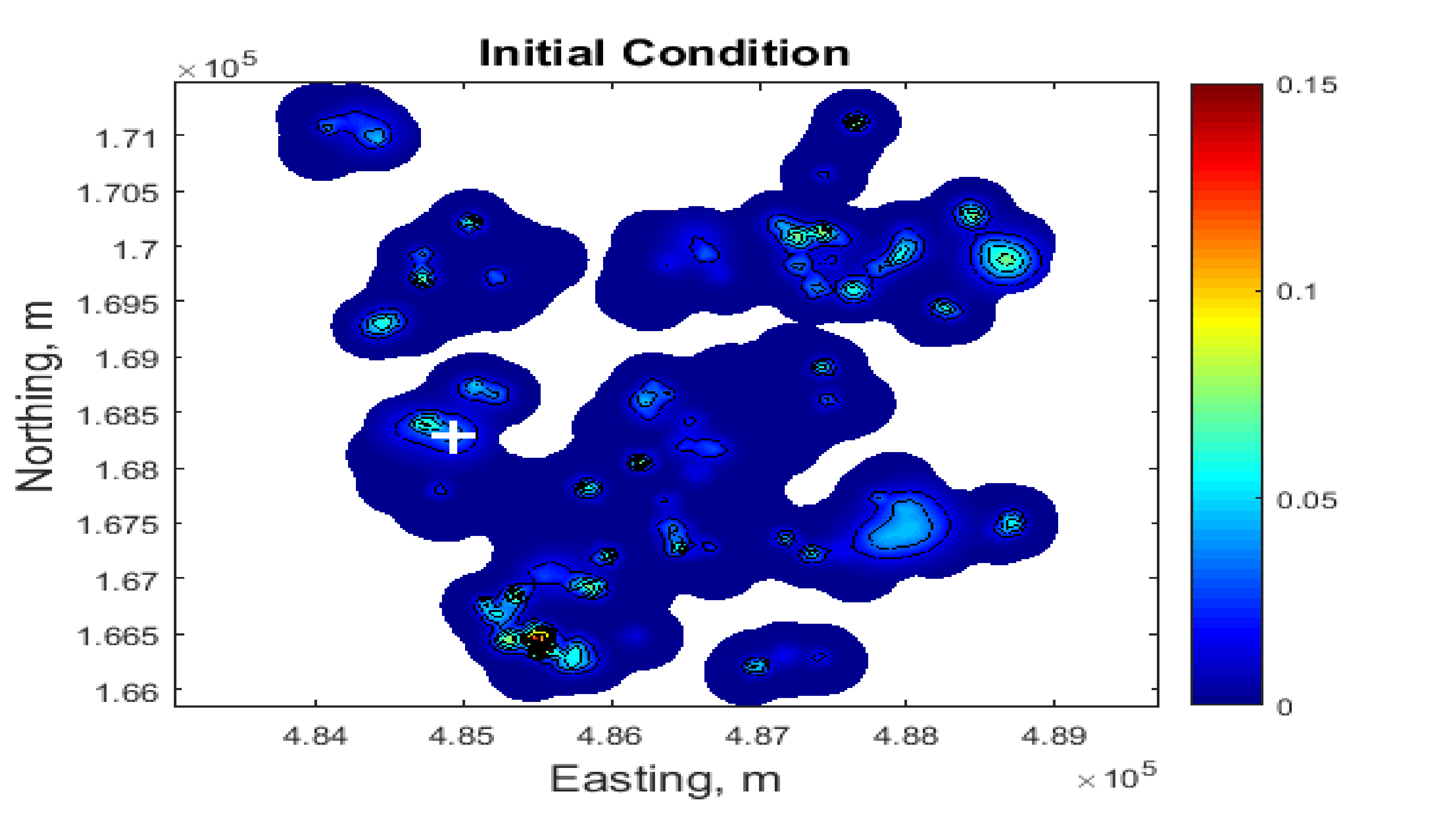}
\includegraphics[width=3.2in,height=2.5in]{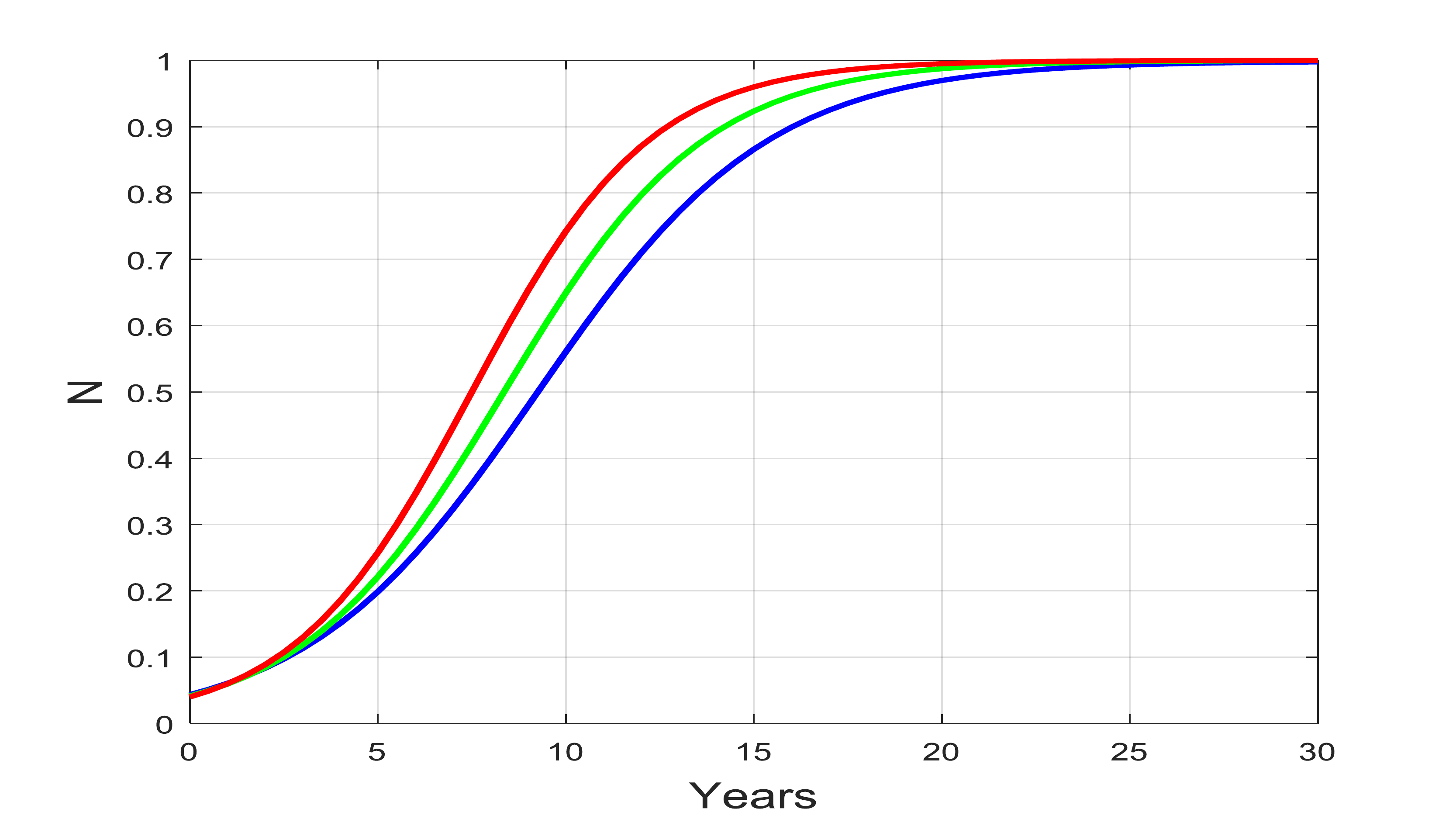}\\
\includegraphics[width=3.2in,height=2.5in]{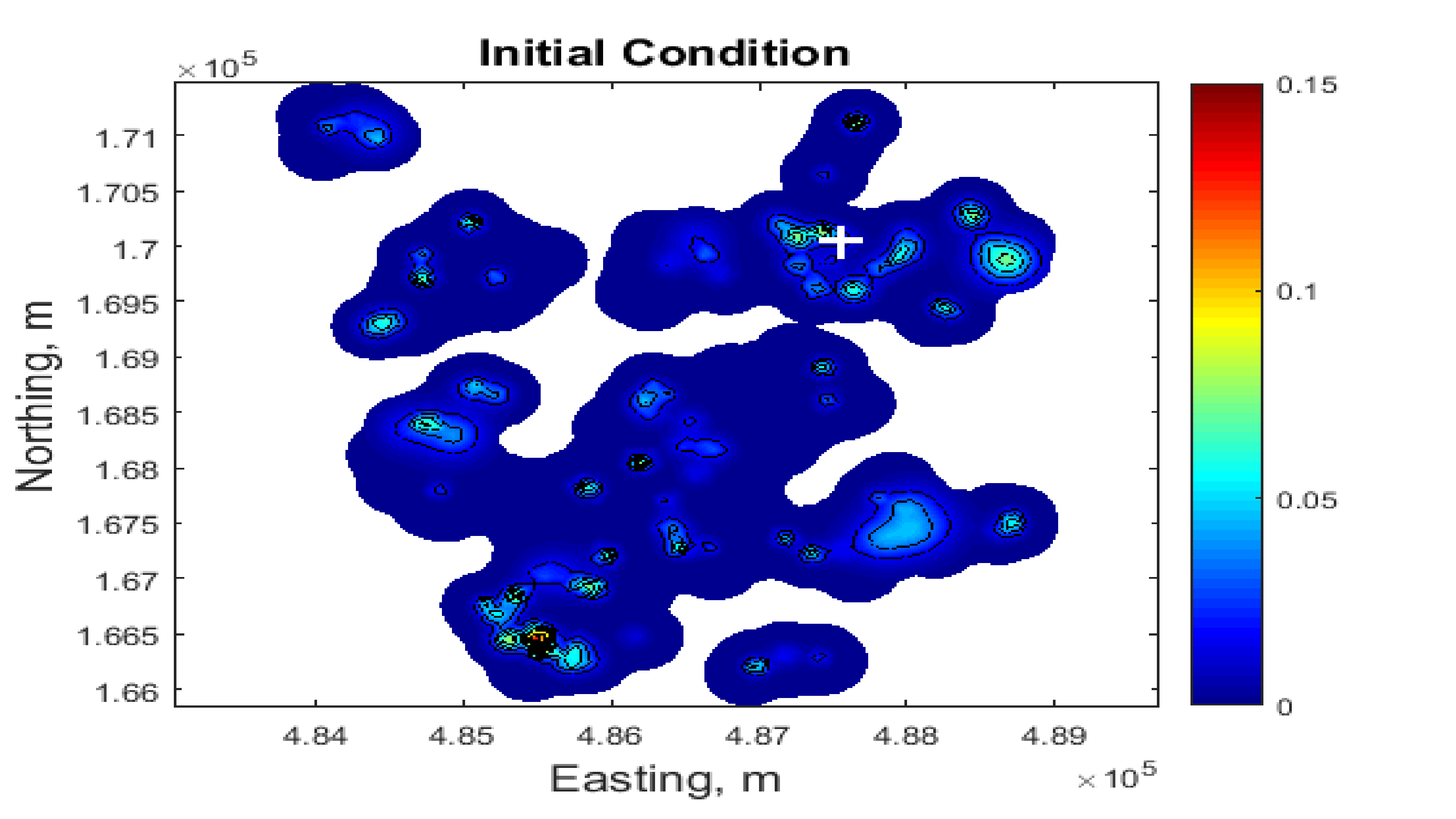}
\includegraphics[width=3.2in,height=2.5in]{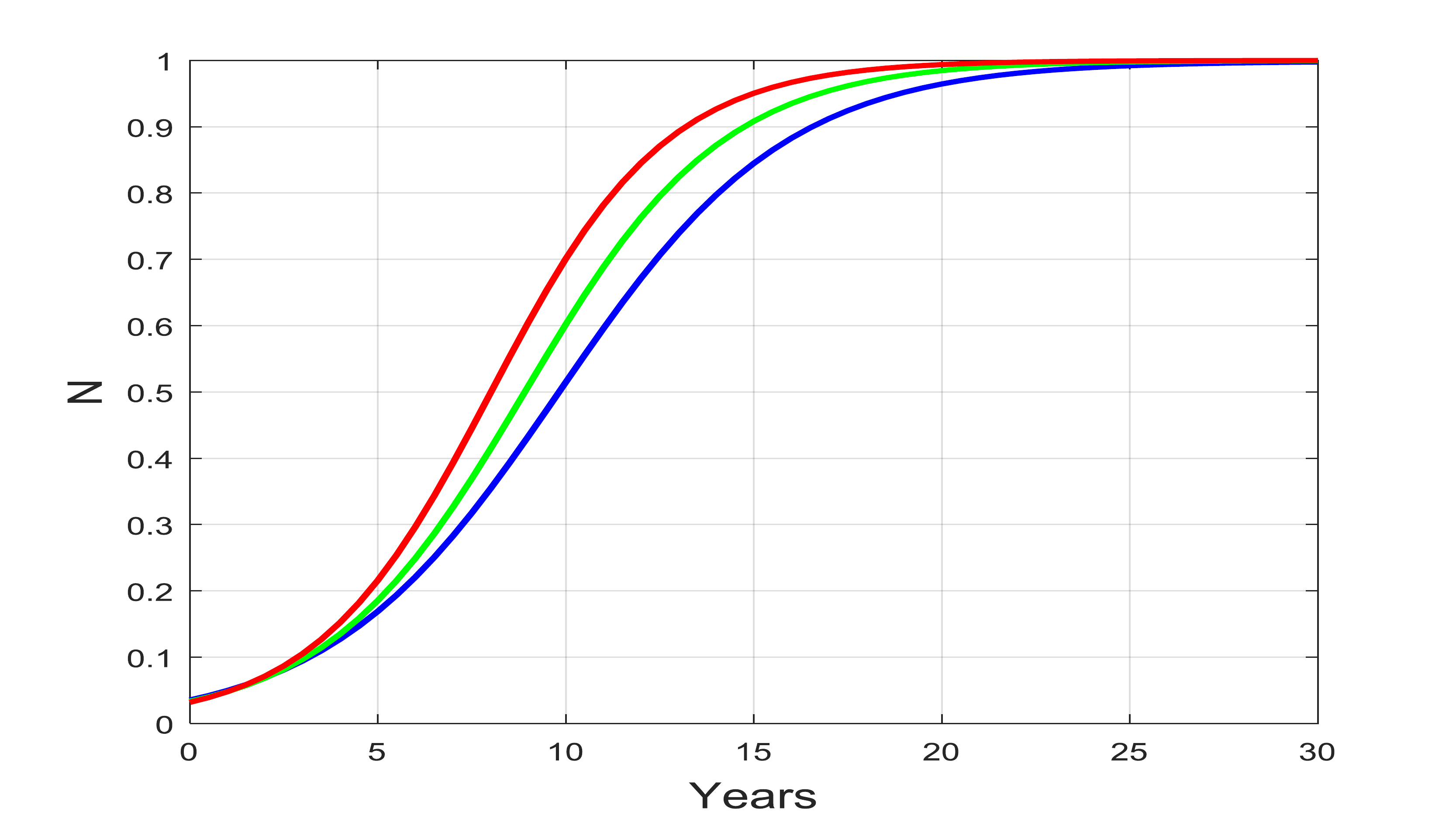}\\
\includegraphics[width=3.2in,height=2.5in]{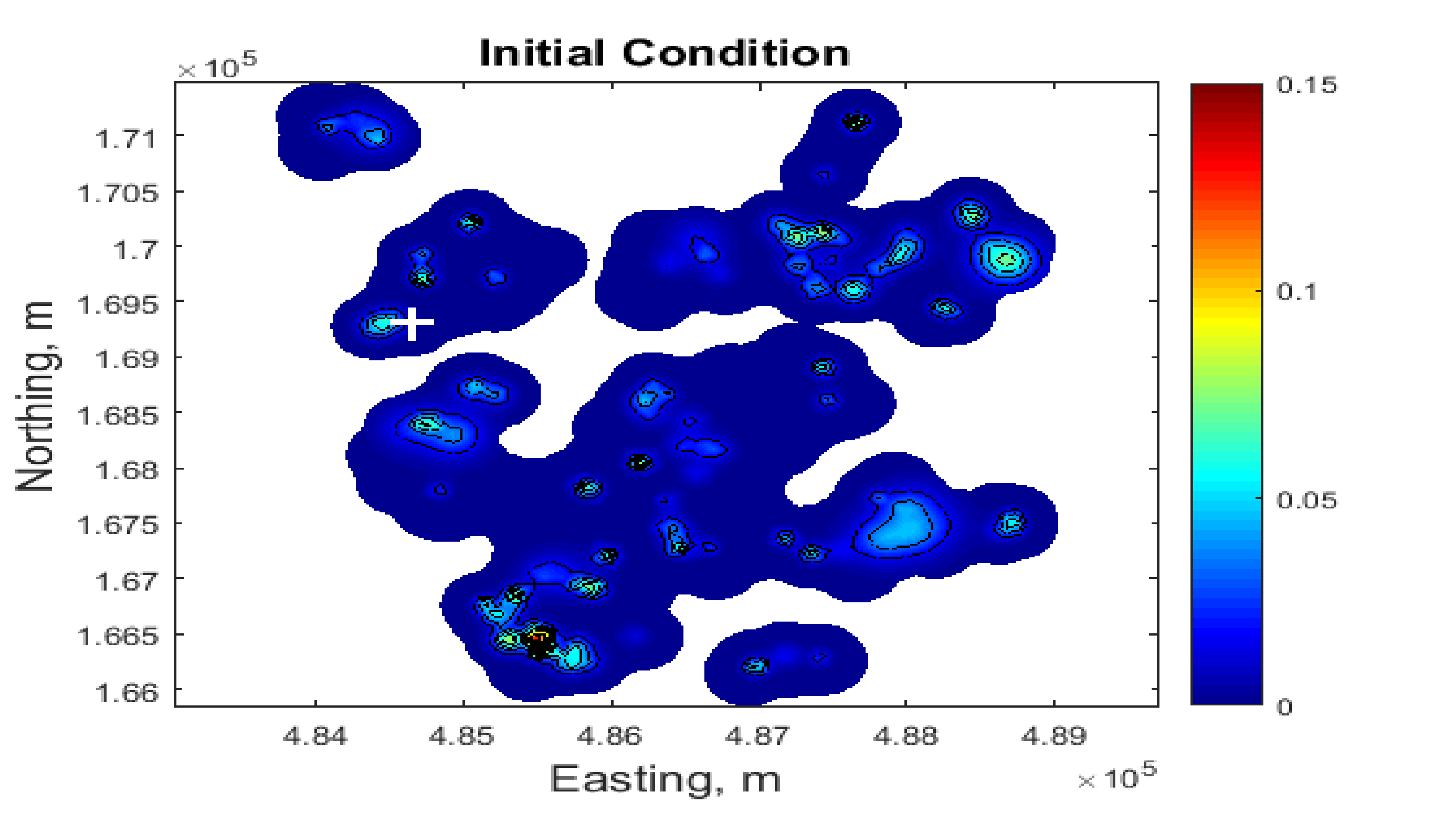}
\includegraphics[width=3.2in,height=2.5in]{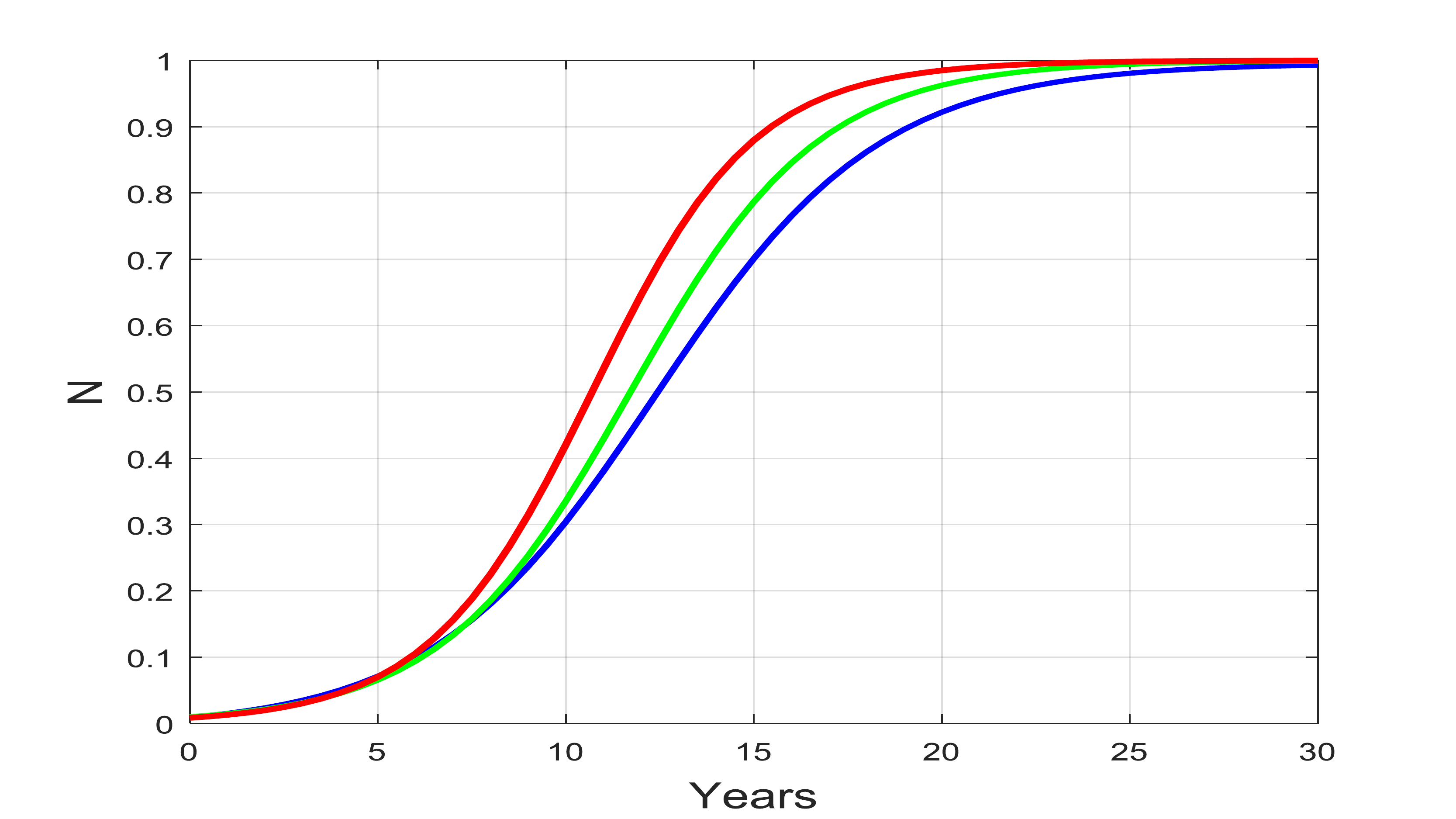}
\caption{Innovation adoption curves at select feeders, where $p=0.001$, $q=0.325$ (blue), $p=0.0001$, $q=0.376$ (green) and $p=0.00001$, $q=0.425$ (red). Top: Feeder location $(484930, 168300)$. Middle: Feeder location $(487560, 170060)$. Bottom: Feeder location $(484650, 169310)$. Left: Feeder location indicated by the white cross on the grid. Right: Innovation adoption curves.
\label{feeders}}
\end{center}
\end{figure}

Next, comparisons of uptake trajectories at different fixed feeder locations are given in Figure \ref{feeders}. On the left, the positioning of the feeder is indicated by a white cross on the grid, and on the right, the adoption ratio as a function of time is displayed for all three cases. Technology clusters form here since these highlighted feeders have non-zero initial conditions, as well, they are influenced by neighbouring feeders with sources. The adoption curves reveal that the rate of uptake at these particular feeders is greater when the magnitude of $p$ decreases and $q$ increases. The exhibited behaviour is again expected since $q$ promotes the spread of technologies at a local level.


\subsection{Uptake at a household level}
We have predicted adoption trajectories at the feeder locations based on the initial condition portrayed by Figure \ref{initcongrid}, right. These curves reveal a macroscopic viewpoint of the adoption behaviour over time, although, the adoption activity at a household level has not yet been discussed. To estimate this, the feeder curves obtained using the FEM can guide the household selection. More precisely, the number of adoptions at a particular feeder and time is specified by its trajectory (see Figure \ref{feeders}, right for examples of feeder curves).

\begin{figure}
\begin{center}
\includegraphics[width=3.2in,height=2.9in]{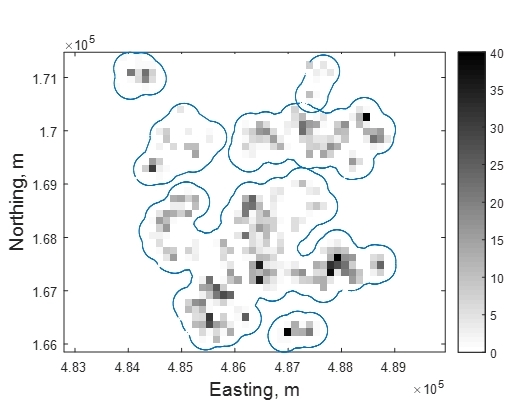}
\includegraphics[width=3.2in,height=2.9in]{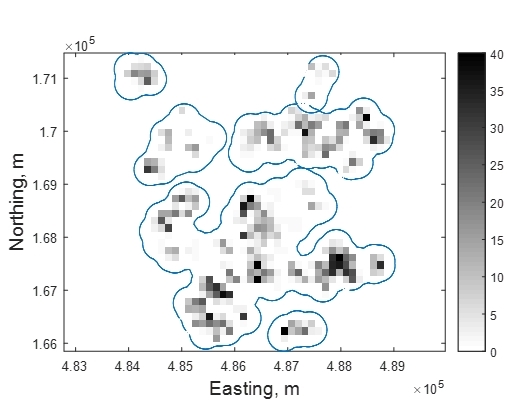}\\
\includegraphics[width=3.2in,height=2.8in]{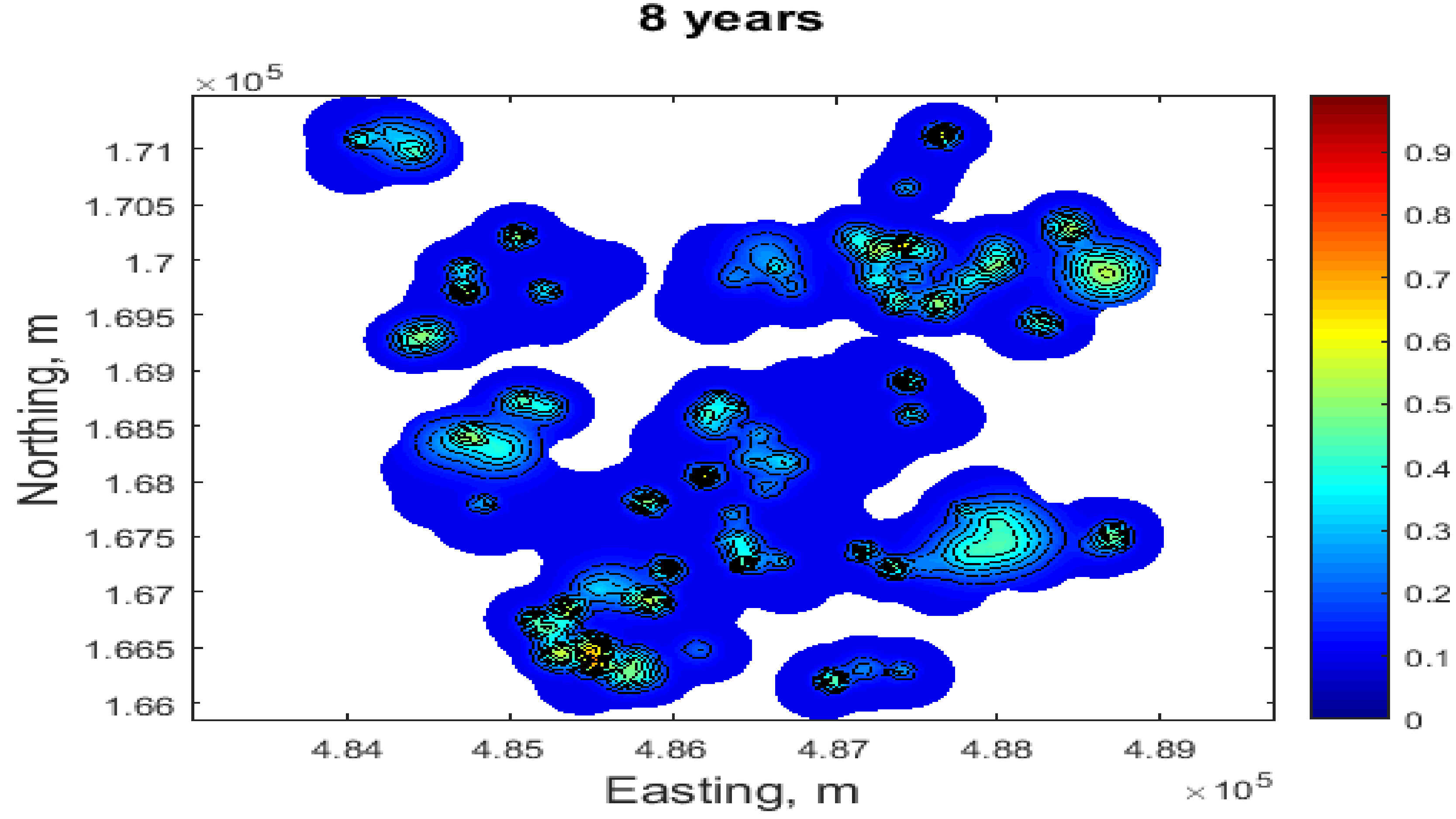}
\includegraphics[width=3.2in,height=2.8in]{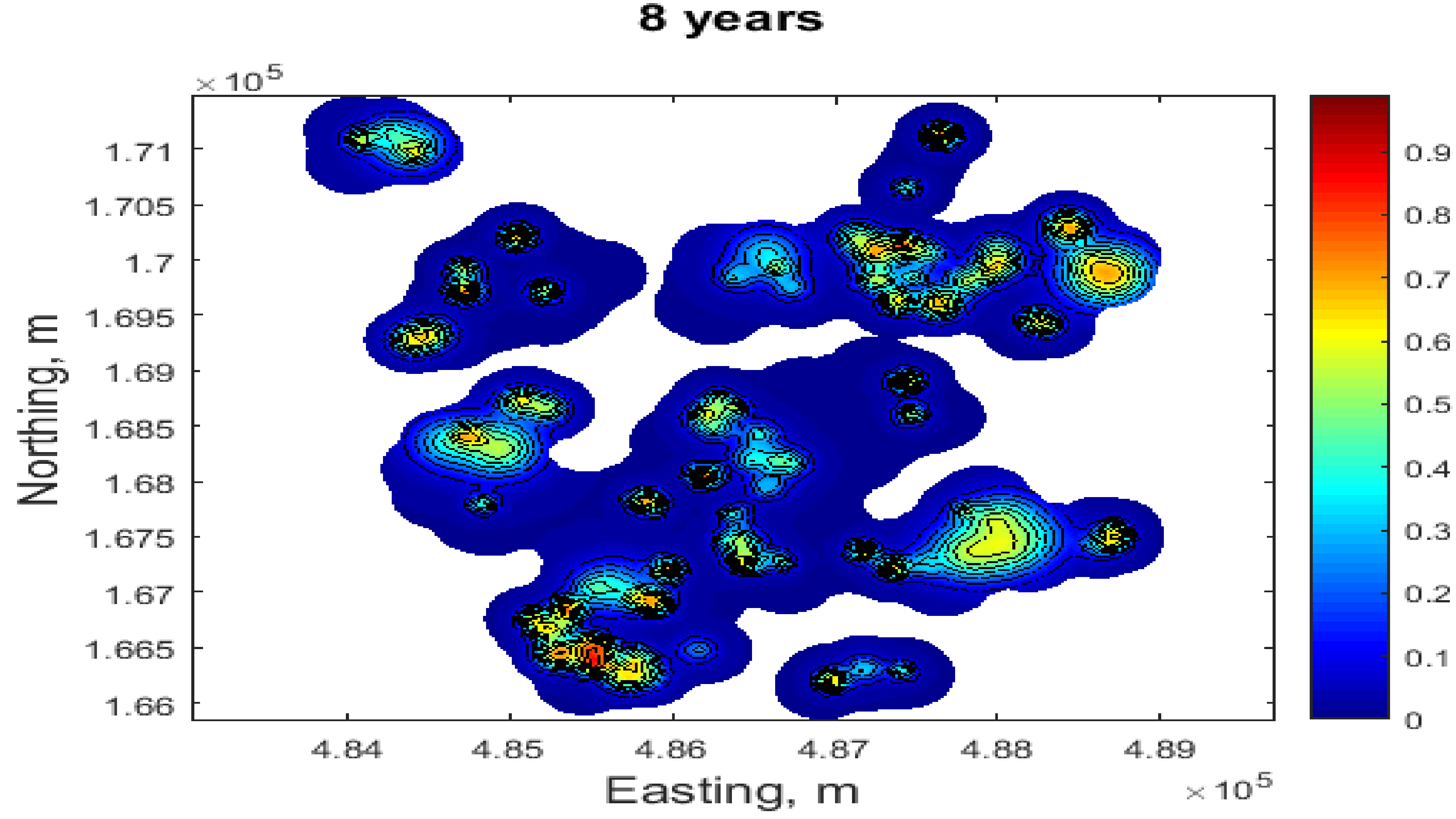}\\
\includegraphics[width=2.6in,height=1.9in]{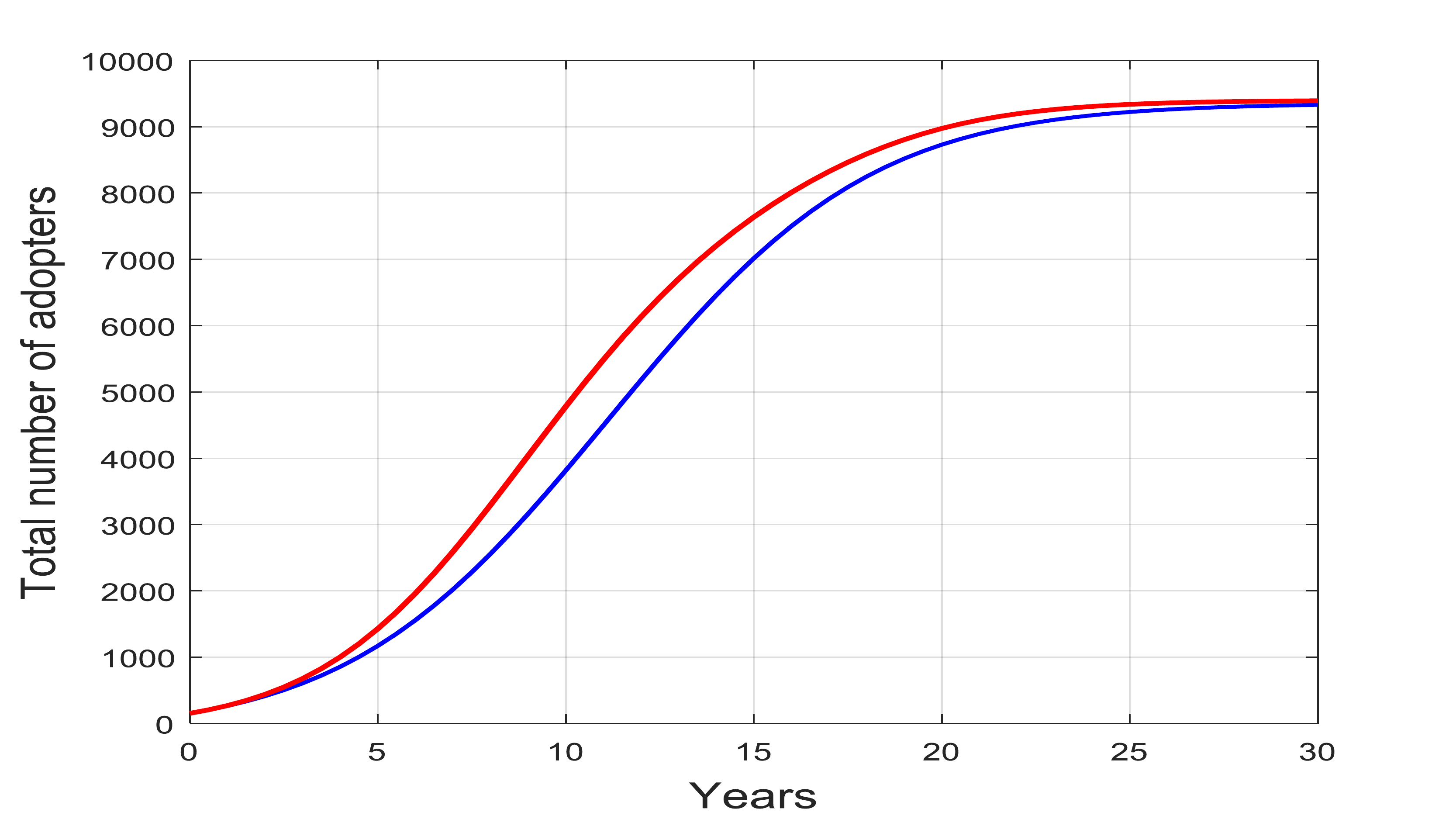}
\caption{Household level uptake. Top Left: Data density plot that represents $2688$ PV households for $p=0.001$, $q=0.325$. The legend indicates the number of PV households at this location. Top Right: Data density plot that represents $3382$ PV households for $p=0.00001$, $q=0.425$. The legend indicates the number of PV households at this location. Middle: Innovation diffusion maps at $t=8$ years, for $p=0.001$, $q=0.325$ (left) and for $p=0.00001$, $q=0.425$ (right). The legend indicates the adoption ratio, $N$. Bottom: Overall number of adopters as a function of time, where the red and blue curve signify $p=0.00001$, $q=0.425$ and $p=0.001$, $q=0.325$ respectively.\label{y8hh}}
\end{center}
\end{figure}

An example of household uptake after eight years is depicted for case $(a)$ and $(c)$ in Figure \ref{y8hh}, where the properties were randomly selected such that the number of adoptions reflect the FEM results at $t=8$ years. The top and middle panels reveal PV household density plots and the FEM results respectively. As well, the overall adoption numbers are given as a function of time along the bottom panel of Figure \ref{y8hh}. A more sophisticated technique can be applied to assign technologies at a household level by using an agent-based approach, where certain households are favoured due to individual characteristics. Refer to \citet{hat17} where an agent-based model was applied to a subset of the geographic area examined here. This study forecasted electric vehicle and photovoltaic uptake using socio-demographic information to make certain households more likely of LCT adoption.

Upon close inspection of the top panel of Figure \ref{y8hh}, it becomes apparent that more densely populated and compact clusters develop when $p$ is reduced and $q$ is increased, which is expected. Interestingly, the number of household adoptions after eight years is $2688$ and $3382$ for case $(a)$ and case $(c)$ respectively. Moreover, the overall adoption curves reveal that the case $(c)$ parameters encourage increased uptake numbers for approximately $t\in(0,30)$, where both trends converge as $t\rightarrow 30$ years. This can be attributed to the feeders being closely spaced and the intense adoption hot-spots that form as a result of the FEM.

\section{Conclusion}
The diffusion of innovations over time and space was explored using the model (\ref{maine}), which was initially proposed by \citet{hay77}. More specifically, we concentrated on the spread of photovoltaics amongst $9484$ households that were linked to an electricity network within Bracknell, UK comprising of $249$ feeders. All households were allocated to a particular feeder, where the feeder co-ordinates were defined as the midpoint of all its connected properties. These co-ordinates then allowed us to generate a grid that represented the local network under investigation. Once this solution domain was outlined, the FEM was utilised to forecast PV uptake with (\ref{maine}), which was dependent upon the coefficients $p$ (innovation) and $q$ (imitation). By then relating these parameters to the well-known Fisher travelling wave solution derived by \citet{abl79}, estimates for the adoption times were found as a function of $p$ and $q$. However, since currently not enough information is available to select accurate initial model conditions and imitation/innovation coefficients, various illustrative cases were instead presented. Here, differing values of $p$ and $q$ were applied that corresponded to a consistent adoption time of $\Delta t=30$ years. As a result, the impact of both these factors on our local network was assessed over a set time frame. Next, the results of our simulations were detailed, where more obvious and intense innovation clusters developed over time when the imitation (internal) effects dominated. In addition, feeder uptake curves were depicted, which were projections for the adoption ratio at some feeder as a function of time. These measures revealed that adoption rates rose at feeders with sources if the imitation influences were increased. Moreover, the feeder curves provided an overall, aggregate understanding of the innovation diffusion process. These projections are potentially very informative estimates for network planners since the predicted increase in load can be accounted for now so to avoid future network issues. Lastly, modelling technology uptake at a household level was addressed, where houses were randomly assigned PVs by following the feeder trends, combining both macroscopic and microscopic approaches. The results presented here suggested that more households installed a PV if the imitation effect outweighed innovation considerably, which was due mainly to the closely spaced feeder locations. Thus, a model has been outlined to estimate the spread of technologies and the impact this will have on a local electricity network, where internal and external influences were considered. When further information becomes available so that the calibration can be improved, then this model will become an important predictive tool for network planners.
\section*{Acknowledgment} {This work was carried out with the support of Scottish and Southern
Electricity Networks through the New Thames Valley Vision
Project (SSET203 New Thames Valley Vision) funded by the Low
Carbon Network Fund established by Ofgem.}
\bibliographystyle{plainnat2}
\bibliography{bib}

\end{document}